\documentclass[useAMS,usenatbib,referee,psfig,epsf]{mn2e}
\usepackage{times,psfig}

\title[Observations of pulsating subdwarf B stars]
{Resolving the pulsations of subdwarf B stars: HS~0039+4302,
HS~0444+0458, and an examination of the group properties of
resolved pulsators}

\author[M.D. Reed et al.]{
 M. D. Reed,$^1$\thanks{E-mail:MikeReed@missouristate.edu}
 D. M. Terndrup,$^2$ A.-Y. Zhou,$^1$
 \newauthor C. T. Unterborn,$^2$ D. An,$^2$ and J. R. Eggen$^1$ \\
 $^1$Department of Physics, Astronomy and Materials Science,
 Missouri State University, 901 S. National, Springfield, MO 65897 USA \\
 $^2$Department of Astronomy, The Ohio State University, 140 W.
 18th Ave., Columbus, OH 43210 USA}

\date{Accepted
      Received }

\begin{document}

\maketitle

\begin{abstract}
We continue our program of single-site observations of pulsating
subdwarf B (sdB) stars and present the results of extensive time
series photometry of HS~0039+4302 and HS~0444+0458. Both were
observed at MDM Observatory during the fall of 2005. We extend the
number of known frequencies for HS~0039+4302 from 4 to 14 and
discover one additional frequency for HS~0444+0458, bringing the
total to three. We perform standard tests to search for multiplet
structure, measure amplitude variations, and examine the frequency
density to constrain the mode degree $\ell$.

Including the two stars in this paper, 23 pulsating sdB stars have
received follow-up observations designed to decipher their
pulsation spectra. It is worth an examination of what has been
detected. We compare and contrast the frequency content in terms
of richness and range and the amplitudes with regards to
variability and diversity. We use this information to examine
observational correlations with the proposed $\kappa$ pulsation
mechanism as well as alternative theories.
\end{abstract}


\section{Introduction}

Subdwarf B (sdB) stars are horizontal-branch stars with masses
$\approx 0.5 M_\odot$, thin ($< 10^{-2} M_\odot$) hydrogen shells,
and temperatures from $22\,000$ to $40\,000$~K
\citep{heber,saf94}. Pulsating sdB stars come in two varieties:
short period (90 to 600 seconds), and long period (45 minutes to 2
hours). This work concentrates on the short-period pulsators,
which are named EC~14026-2647 stars after the prototype
\citep{kill97}; they are also known as V361~Hya stars or sdBV
stars. They typically have pulsation amplitudes near 1\%, and
detailed studies reveal a few to dozens of frequencies. The longer
period pulsators are known as PG~1716 pulsators after that prototype
and typically have amplitudes less then 0.1\% \citep{grn03}. They are also 
cooler then the EC~14026-type pulsators, though there is some overlap, and
they are most likely $g-$mode pulsators \citep{font}.

Asteroseismology of pulsating sdB stars can potentially probe the
interior structure and provide estimates of total mass, shell
mass, luminosity, helium fusion cross sections, and coefficients
for radiative levitation and gravitational diffusion. To apply the
tools of asteroseismology, however, it is necessary to resolve the
pulsation frequencies.  This usually requires extensive
photometric campaigns, preferably at several sites spaced in
longitude to reduce day/night aliasing. Generally, discovery
surveys have simply identified variables and detected only the
highest-amplitude pulsations, while multisite campaigns have
observed few sdBV stars.

We have been engaged in a long-term program to resolve
poorly-studied sdB pulsators, principally from single-site data.
This method has proven useful for several sdBV stars
\citep{me2,reed06a,reed07,zhou}.  Here, we report the results of
our observations of HS~0039+4302 (hereafter HS~0039) and
HS~0444+0458 (hereafter HS~0444). HS~0039 ($B = 15.6$) and HS~0444
($B = 16.5$) were discovered to be members of the EC~14026 class
by \citet[][hereafter {{\O}01}]{ost01a}. Their observations of
HS~0039 consisted of three runs of 1, 2, and 3 hours, and they
obtained three $\approx 1.5$ hour runs for HS~0444.  Even with
these short runs, they were able to detect four frequencies in
HS~0039 and two in HS~0444. {\O}01 also obtained spectra of both
stars, from which they determined $T_{\rm eff} = 33~100$ K and
$34~500$ K, and $\log g = 6.0$ and $6.1$ 
(with $g$ in cgs units of $cm\cdot s^{-2}$) for HS~0039 and HS~0444,
respectively, and that neither star is a spectroscopic binary. In
\S 2 we describe our new observations of these stars, in \S 3 we
analyze the pulsation frequencies, and in \S 4 we discuss our
findings and apply asteroseismic tests.

With the addition of these two stars, 23 sdBV stars have received
follow-up observations, including 18 stars for which our program
has directly contributed data. In \S 5, we will discuss the
observational properties of all 23 stars, concentrating on
pulsation stability, amplitudes, and a comparison with known
driving mechanisms.

\section{Observations}
Data were obtained at MDM Observatory's 1.3~m telescope using an
Apogee Alta U47+ CCD camera. MDM Observatory is located on the southwest
ridge of Kitt Peak, Arizona and is operated by a consortium of five
universities, including the Ohio State University.  
Images were transferred via USB2.0
for high-speed readout; our binned ($2\times 2$) images had an
average dead-time of one second. The observations used a red
cut-off filter (BG38), so the effective bandpass covers the $B$
and $V$ filters and is essentially that of a blue-sensitive
photomultiplier tube. Such a setup allows us to maximize light
throughput while maintaining compatibility with observations
obtained with photomultipliers.  Tables~\ref{tab01} and
\ref{tab02} provide the details of our observations including
date, start time, run length, and integration time. The
observations total nearly 100 hours of data for HS~0039 and more
than 60 hours for HS~0444.

\begin{figure}
 \psfig{figure=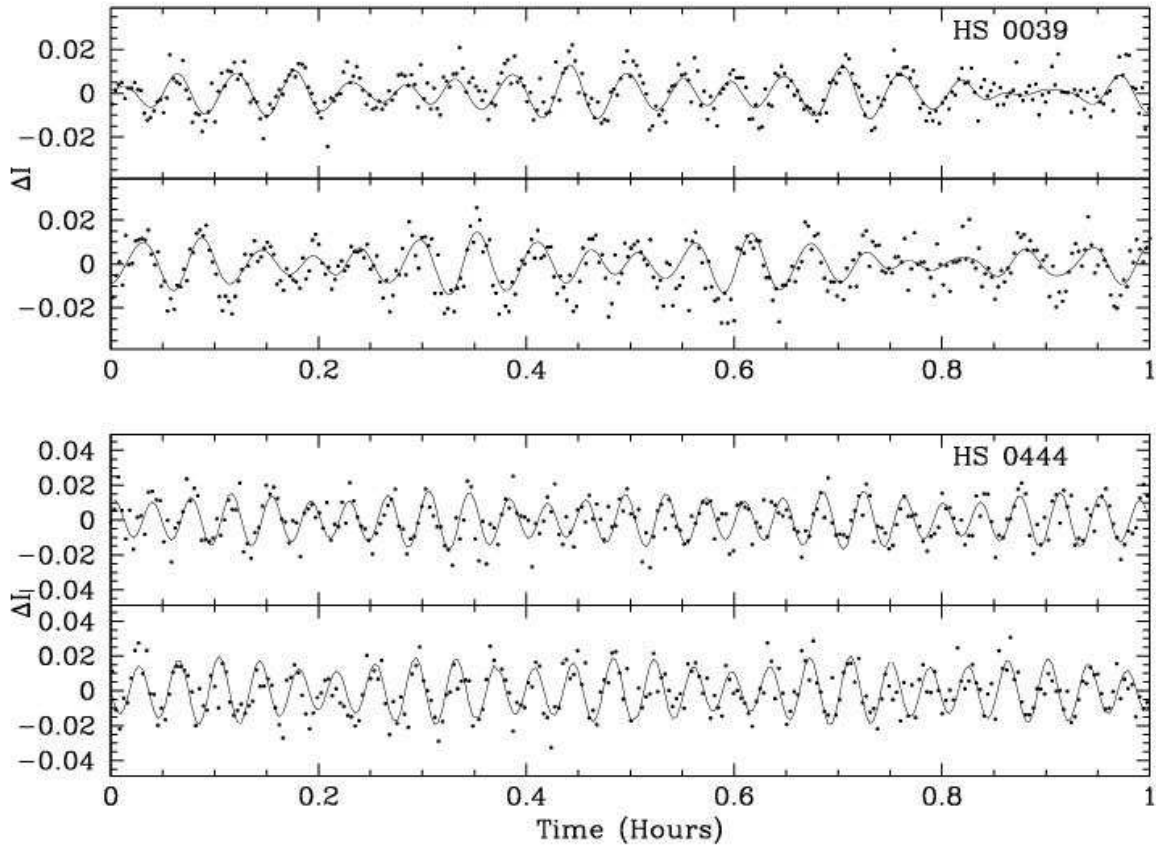,angle=-90,width=\textwidth}
\caption{Representative lightcurves of HS~0039 (top pair) and
HS~0444 (bottom pair). Each panel is one hour in length. The line
shows our fit to the data.  Note that the vertical scale differs
for the two plots.} \label{fig01}
\end{figure}

Standard image reduction procedures, including bias subtraction,
dark current and flat field correction, were followed using
IRAF\footnote{IRAF is distributed by the National Optical
Astronomy Observatories, which are operated by the Association of
Universities for Research in Astronomy, Inc., under cooperative
agreement with the National Science Foundation.} packages.
Intensities were extracted using aperture photometry.  Extinction
and cloud corrections were obtained from the normalized
intensities of several field stars.  Because sdB stars are
substantially hotter than typical field stars, differential light
curves are not flat due to atmospheric reddening. A low-order
polynomial was fit to remove nightly trends from the data.
Finally, the lightcurves were normalized by their average flux and
centered around zero so the reported differential intensities are
$\Delta I = \left( I /\langle I\rangle\right) -1 $. Amplitudes are
given as milli-modulation amplitudes (mma), with 10~mma
corresponding to 1.0\% or 9.2~millimagnitudes. Sample lightcurves
are shown in Fig.~\ref{fig01}.

%Table 1
\begin{table}
 \caption{Observations of HS~0039 \label{tab01}}
 \begin{tabular}{@{}lcccc} \hline
Run & UT Start & UT Date & Length & Integration \\
 & (h:m:s) & 2005 & (hr) & (s)\\ \hline
 mdm111505 & 01:46:00 & 15 Nov. & 7.9 & 15 \\
 mdm111605 & 01:17:00 & 16 Nov. & 8.4 & 12 \\
 mdm111705 & 01:18:00 & 17 Nov. & 8.4 & 12 \\
 mdm111805 & 01:16:00 & 18 Nov. & 8.5 & 12 \\
 mdm111905 & 01:15:00 & 19 Nov. & 8.5 & 12 \\
 mdm112005 & 01:23:00 & 20 Nov. & 8.3 & 12 \\
 mdm112105 & 01:23:00 & 21 Nov. & 8.2 & 10 \\
 mdm112205 & 01:11:00 & 22 Nov. & 8.6 & 10 \\
 mdm112505 & 02:35:00 & 25 Nov. & 6.5 & 12 \\
 mdm112605 & 01:06:00 & 26 Nov. & 4.0 & 12 \\
 mdm112705 & 01:11:00 & 27 Nov. & 0.6 & 12 \\
 mdm112805 & 01:11:00 & 28 Nov. & 6.2 & 10 \\
 mdm112905 & 01:06:00 & 29 Nov. & 6.4 & 10 \\
 mdm113005 & 01:90:00 & 30 Nov. & 3.4 & 10 \\
 mdm120905 & 01:24:00 & 09 Dec. & 4.7 &  8 \\
 mdm121405 & 01:14:00 & 14 Dec. & 1.1 &  8 \\ \hline
\end{tabular}
\end{table}

%Table 2
\begin{table}
 \caption{Observations of HS~0444 \label{tab02}}
 \begin{tabular}{@{}lcccc} \hline
 Run & UT Start & UT Date & Length & Integration \\
 & (h:m:s) & 2005 & (hr) & (s)\\ \hline
 hs04mdm111505 & 09:54:00 & 15 Nov. & 3.0 & 15 \\
 hs04mdm111605 & 09:52:00 & 16 Nov. & 3.1 & 12 \\
 hs04mdm111705 & 09:53:00 & 17 Nov. & 3.0 & 12 \\
 hs04mdm111805 & 09:52:00 & 18 Nov. & 2.9 & 12 \\
 hs04mdm111905 & 09:53:00 & 19 Nov. & 2.8 & 12 \\
 hs04mdm112005 & 09:57:00 & 20 Nov. & 2.7 & 12 \\
 hs04mdm112105 & 09:48:00 & 21 Nov. & 2.8 & 12 \\
 hs04mdm112205 & 10:52:00 & 22 Nov. & 1.6 & 12 \\
 hs04mdm112605 & 05:26:00 & 26 Nov. & 6.8 & 12 \\
 hs04mdm112705 & 08:59:00 & 27 Nov. & 3.1 & 12 \\
 hs04mdm112805 & 07:23:10 & 28 Nov. & 4.8 & 10 \\
 hs04mdm112905 & 07:32:00 & 29 Nov. & 4.5 & 10 \\
 hs04mdm113005 & 04:26:00 & 30 Nov. & 4.7 & 10 \\
 hs04mdm120905 & 06:26:40 & 09 Dec. & 2.6 & 15 \\
 hs04mdm121005 & 01:30:30 & 10 Dec. & 9.5 & 10 \\
 hs04mdm121405 & 05:45:10 & 14 Dec. & 5.3 & 12 \\ \hline
\end{tabular}
\end{table}

\section{Pulsation Analysis}
\begin{figure}
 \centerline{\psfig{figure=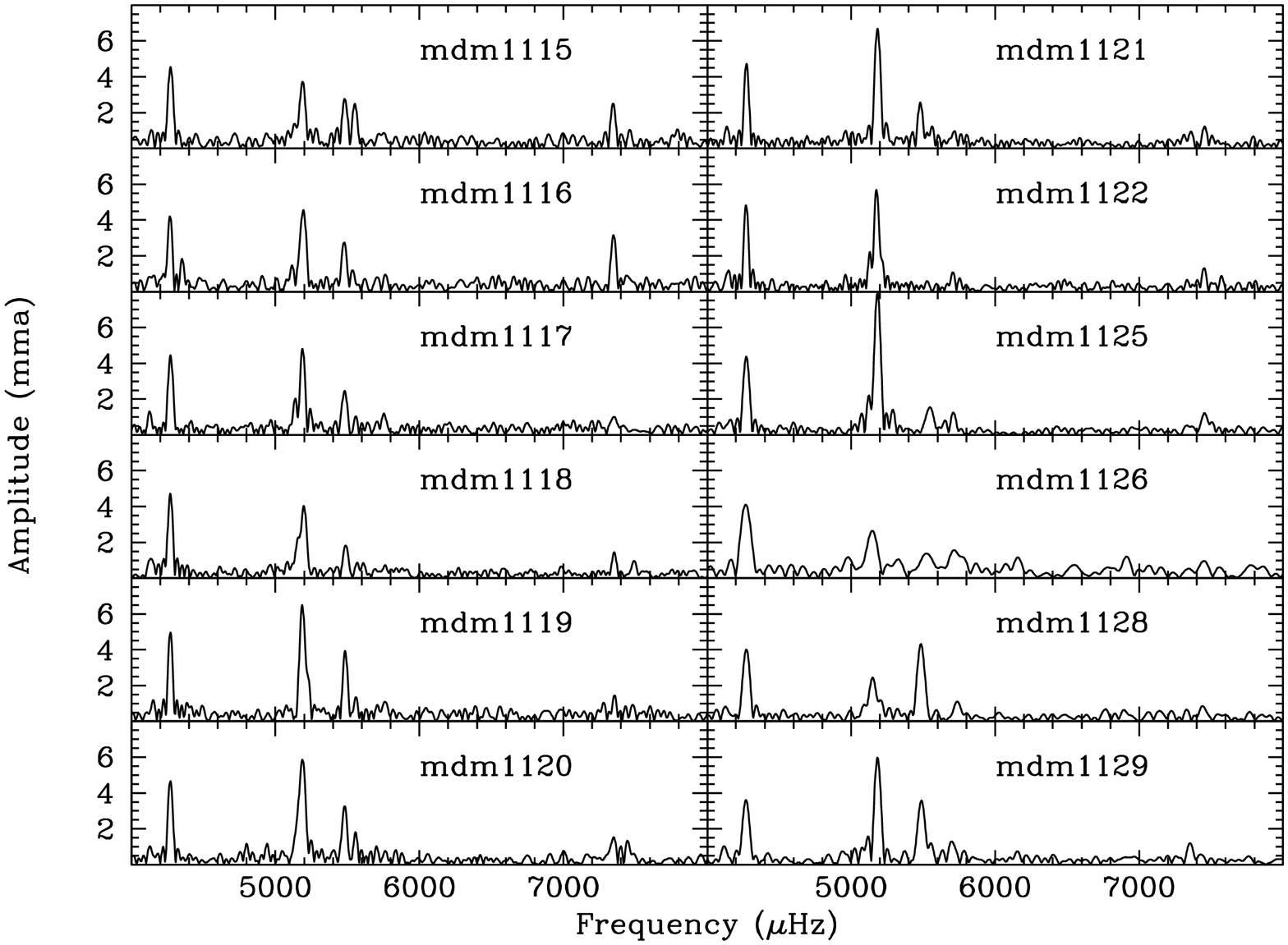,angle=-90,width=\textwidth}}
\caption{Pulsation spectra of temporally adjacent data runs for
HS~0039 plotted at the same scale.} \label{fig02}
\end{figure}

\begin{figure}
 \centerline{\psfig{figure=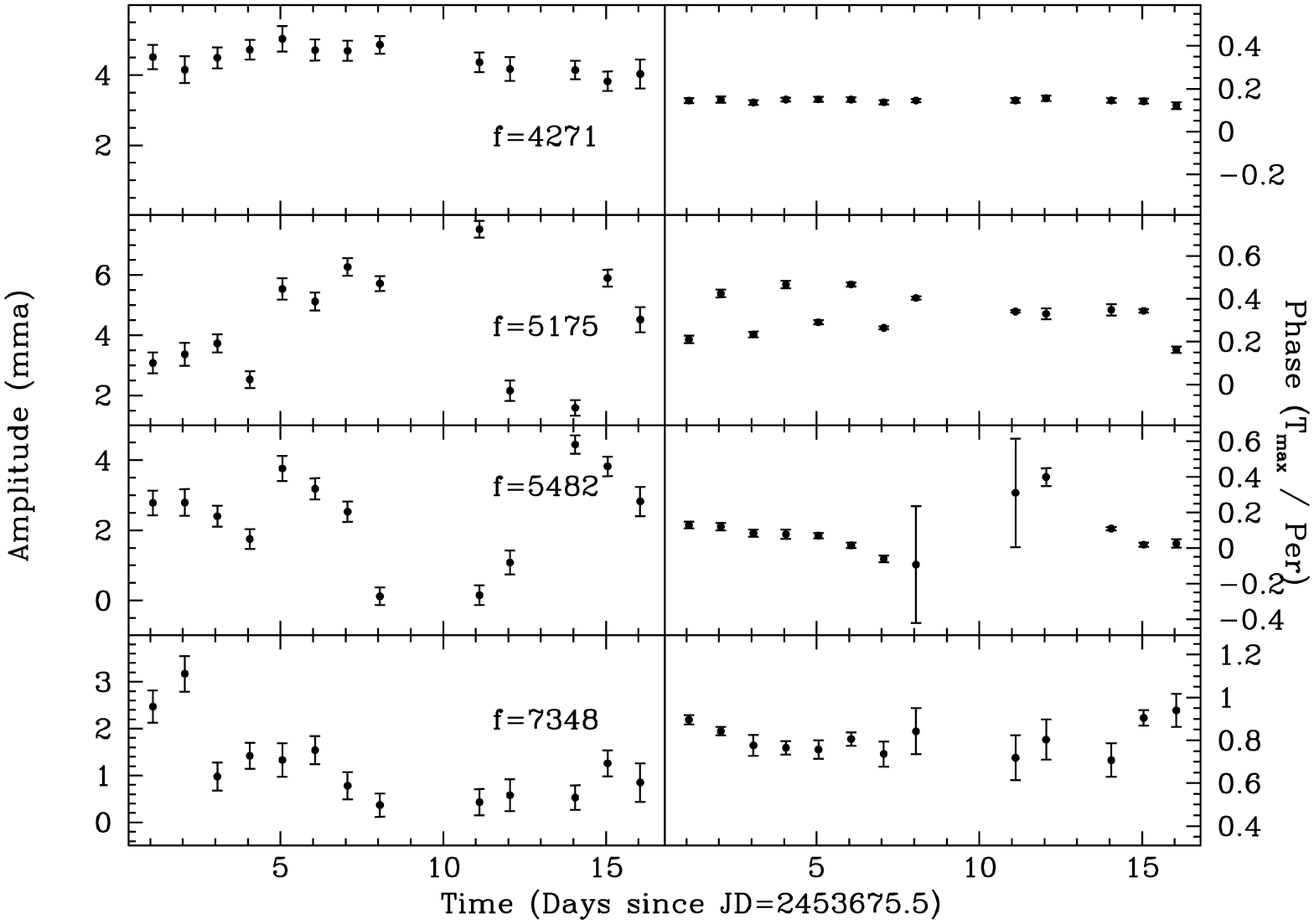,angle=-90,width=\textwidth}}
\caption{Amplitudes and phases of the four largest amplitude
frequencies of HS~0039. Frequencies are provided (in $\mu$Hz) in each
panel.\label{fig03}}
\end{figure}

{\bf HS~0039: } A quick analysis during observations alerted us
that the amplitudes of HS~0039 were not stable. This is easily
seen in the final nightly reductions, twelve of which are shown in
Fig.~\ref{fig02}. In this figure, we show the pulsation spectra
(Fourier transforms; FTs) from adjacent nights, except for the short night on 27
November.  Only the frequency near 4270~$\mu$Hz appears stable in
amplitude, while those near 5175, 5482, and 7348~$\mu$Hz show
substantial variation.  We therefore examined the amplitudes and
phases for indications of closely spaced multiplets, which would
produce roughly sinusoidal amplitude variations and phase changes
near the median amplitude \citep[see][]{dmp}. Figure~\ref{fig03}
shows our non-linear least-squares fits for the four dominant
frequencies.

\begin{figure}
 \centerline{\psfig{figure=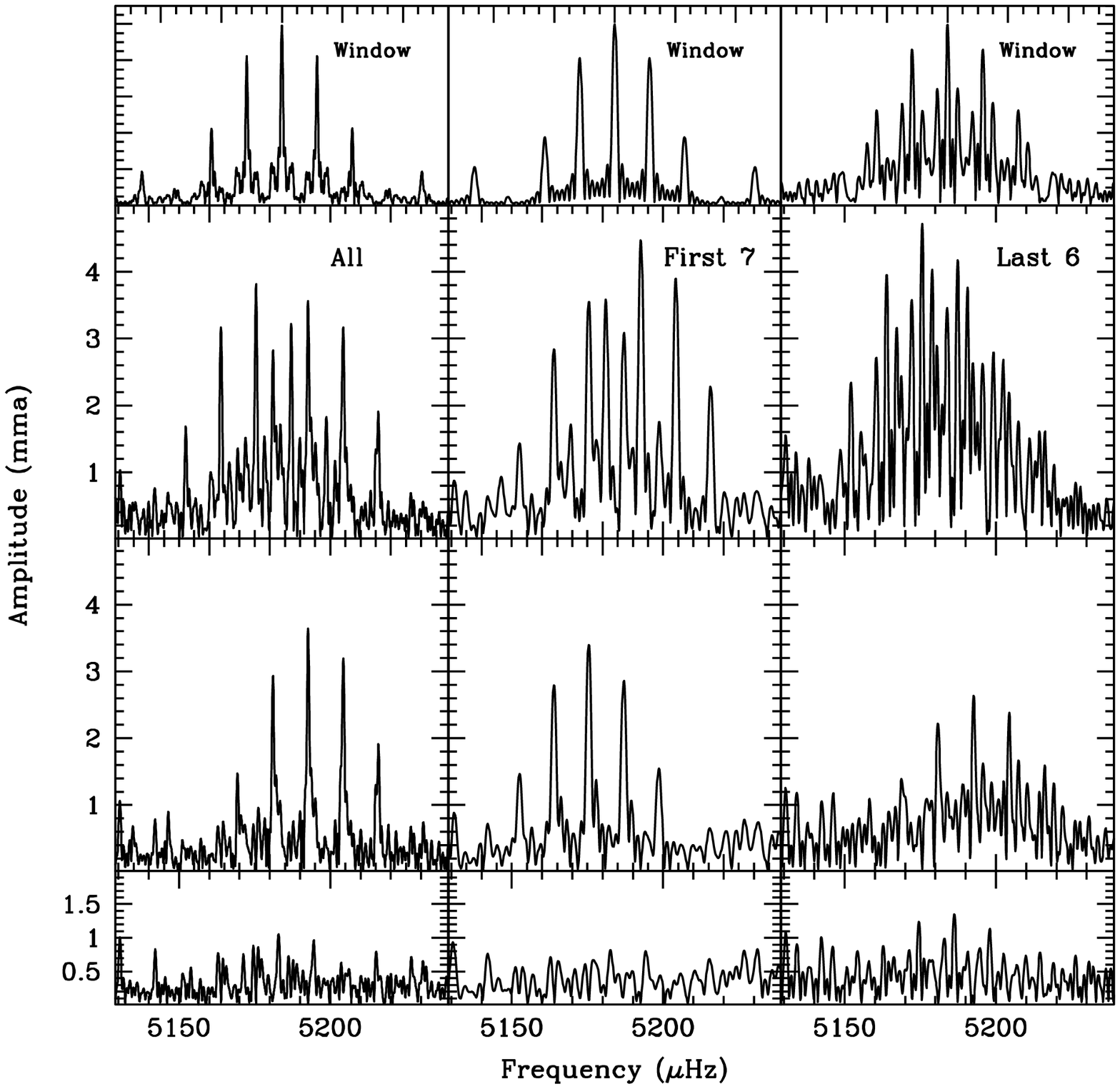,width=\textwidth}}
\caption{Close-up of the region near 5175~$\mu$Hz showing
prewhitening of the frequency doublet in HS~0039. From top to
bottom the panels are the window functions, the original FTs, and
the FTs prewhitened by one, and two frequencies. The left panels
are the complete data, the middle panels are the first seven data
runs combined, and the right panels are the last six data runs
combined.\label{fig04}}
\end{figure}

The frequency near 4270~$\mu$Hz is the most stable, both in
amplitude and phase. The remaining three frequencies show
significant amplitude variation, but only a little variation in
phase. The phase for the peak near 5175 $\mu$Hz shows an unusual
bimodality at the beginning of the campaign, with a steady,
intermediate value at the end. We therefore separated the data
into two subsets composed of data from the first
seven runs (15 -- 21 Nov.) and the last six runs (25 -- 30 Nov.).
Figure~\ref{fig04} shows the region near 5175~$\mu$Hz for all the
November data and the two subsets. The FT of
the first seven runs allowed us to interpret the phase
information:  the frequency at 5175 $\mu$Hz is composed of a close
doublet separated by $\sim 17 \mu$Hz. The separation between frequencies 
is just shorter than 1 day ($11.6$ $\mu$Hz) and so nightly runs will
not resolve these into two separate frequencies.  Evidently the timing was
just right near the beginning of the campaign that the phase switched
between the two frequencies of the doublet on alternate nights.
This was not the case later in the run, although the very last
phase might show that the pattern was reestablishing itself.

The frequency near 5482 $\mu$Hz shows a similar variation in
amplitude to that near 5175 $\mu$Hz, but not the bimodal phase. We
grouped the data into the same subsets as in Fig.~\ref{fig04},
but did not see any clear sign of a close doublet.  If the
amplitude variations were intrinsic to that peak, the discovery of
frequencies by prewhitening would be more complicated.  We
produced an accurate window function as in \citet{reed06a}; the
window function is the FT of a noise-free sinusoidal single
frequency sampled at the same times as the data.  The window
function matched the multi-peaked structure of the temporal
spectra around 5482 $\mu$Hz, leading us to conclude that the
amplitude and phase variations are intrinsic to that frequency and
it is not a closely-spaced multiplet.  We also examined the region
near 7348~$\mu$Hz in the complete data and in the subsets.  There
is no indication of closely spaced multiplets, nor in the the
phases (Fig.~\ref{fig03}).  We conclude that this frequency also
has intrinsic amplitude variations, but is a temporally resolved
frequency.

\begin{figure}
 \centerline{\psfig{figure=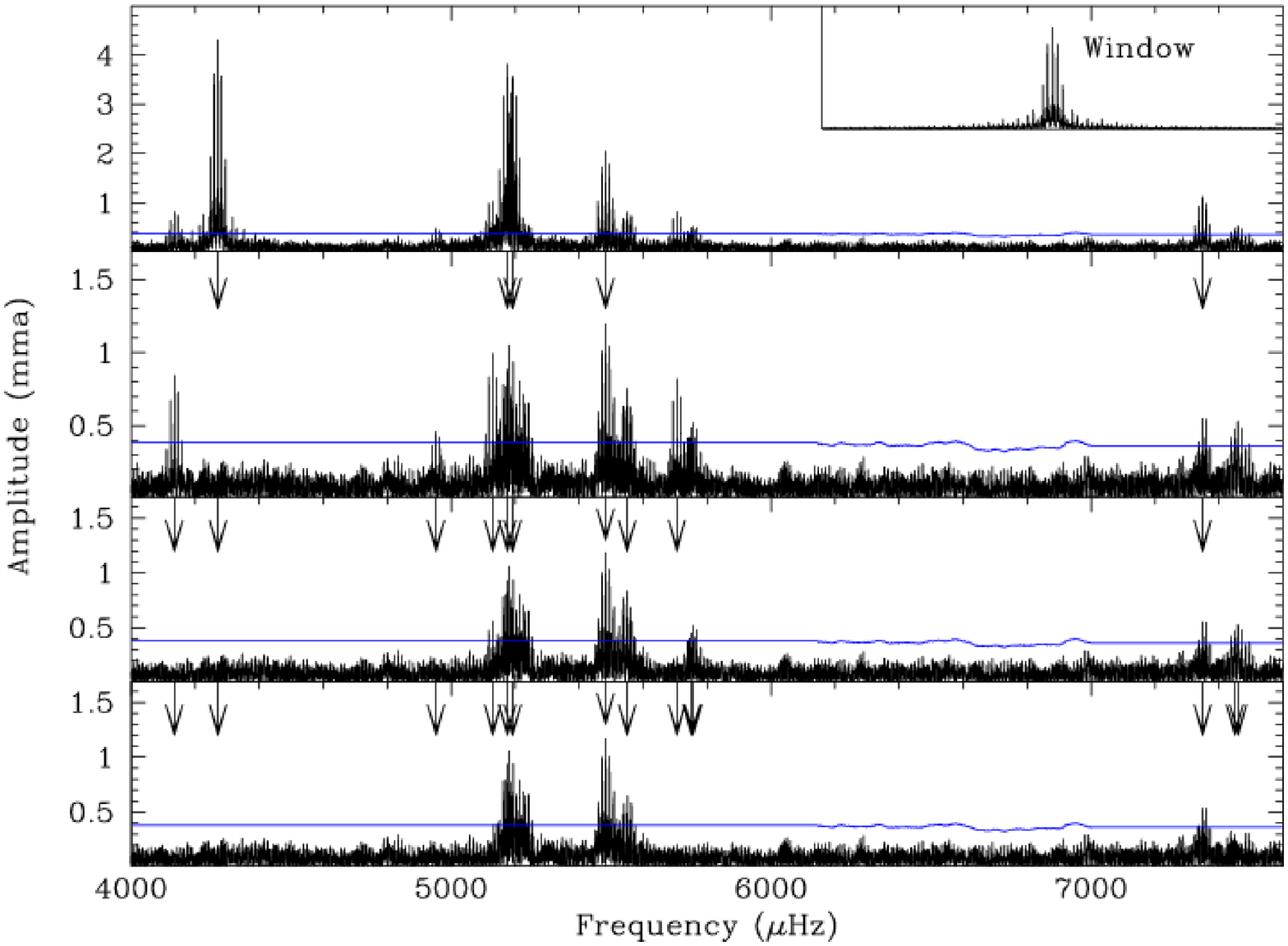,angle=-90,width=\textwidth}}
\caption{Temporal spectrum for the complete HS~0039 data set
showing the prewhitening sequence. The top panel shows the
original FT while the remaining panels show the data prewhitened
by 5, 10, and 14 frequencies, respectively. The prewhitened frequencies
are indicated by arrows and the solid (blue in the
on-line version) lines indicate the $4\sigma$ detection level
while the inset is the window function.\label{fig05}}
\end{figure}

We do not see any such indications in the remaining frequencies,
though most have amplitudes that are too small to be detected in
individual runs. Armed with five resolved frequencies, we
continued simultaneously least-squares fitting and removing peaks
(prewhitening) in the combined data set. This process is shown in
Fig.~\ref{fig05}, where the top panel is the original FT and the
next three panels show residuals after prewhitening by 5, 10, and
14 frequencies (from top to bottom). The solid (blue in the
on-line version) line in the figure indicates the $4\sigma$ noise
limit, below which we do not fit any peaks.  After fitting 14
frequencies, we concluded that any remaining power in the residuals was
caused by amplitude and/or phase variations and so does not
represent remaining unfit frequencies. The solution to our fit,
listing frequencies, periods, and average amplitudes is provided
in Table~\ref{tab03}.

% Table 3
\begin{table}
\caption{Frequencies, periods, and average amplitudes for HS~0039.
Formal least-squares errors are in parentheses.\label{tab03}}
\label{tab03}
\begin{tabular}{lccc}
\hline
ID & Frequency & Period  & Amplitude \\
 & ($\mu$Hz) & (s) & (mma) \\ \hline
$f1$  & 4135.559 (0.042) & 241.8052 (0.0024) & 0.85 (0.09) \\
$f2$  & 4271.481 (0.008) & 234.1108 (0.0004) & 4.35 (0.09) \\
$f3$  & 4952.858 (0.078) & 201.9035 (0.0031) & 0.46 (0.09) \\
$f4$  & 5130.474 (0.036) & 194.9137 (0.0013) & 1.00 (0.09) \\
$f5$  & 5175.466 (0.009) & 193.2193 (0.0003) & 3.99 (0.09) \\
$f6$  & 5192.660 (0.010) & 192.5794 (0.0003) & 3.73 (0.09) \\
$f7$  & 5482.319 (0.016) & 182.4045 (0.0005) & 2.16 (0.09) \\
$f8$  & 5550.170 (0.047) & 180.1746 (0.0015) & 0.77 (0.09) \\
$f9$  & 5705.677 (0.044) & 175.2640 (0.0013) & 0.82 (0.09) \\
$f10$ & 5751.316 (0.081) & 173.8732 (0.0024) & 0.45 (0.09) \\
$f11$ & 5756.651 (0.068) & 173.7120 (0.0020) & 0.53 (0.09) \\
$f12$ & 7348.444 (0.031) & 136.0832 (0.0006) & 1.15 (0.09) \\
$f13$ & 7449.256 (0.064) & 134.2415 (0.0011) & 0.65 (0.09) \\
$f14$ & 7459.765 (0.063) & 134.0524 (0.0011) & 0.67 (0.09) \\
\hline
\end{tabular}
\end{table}

\begin{figure}
 \centerline{\psfig{figure=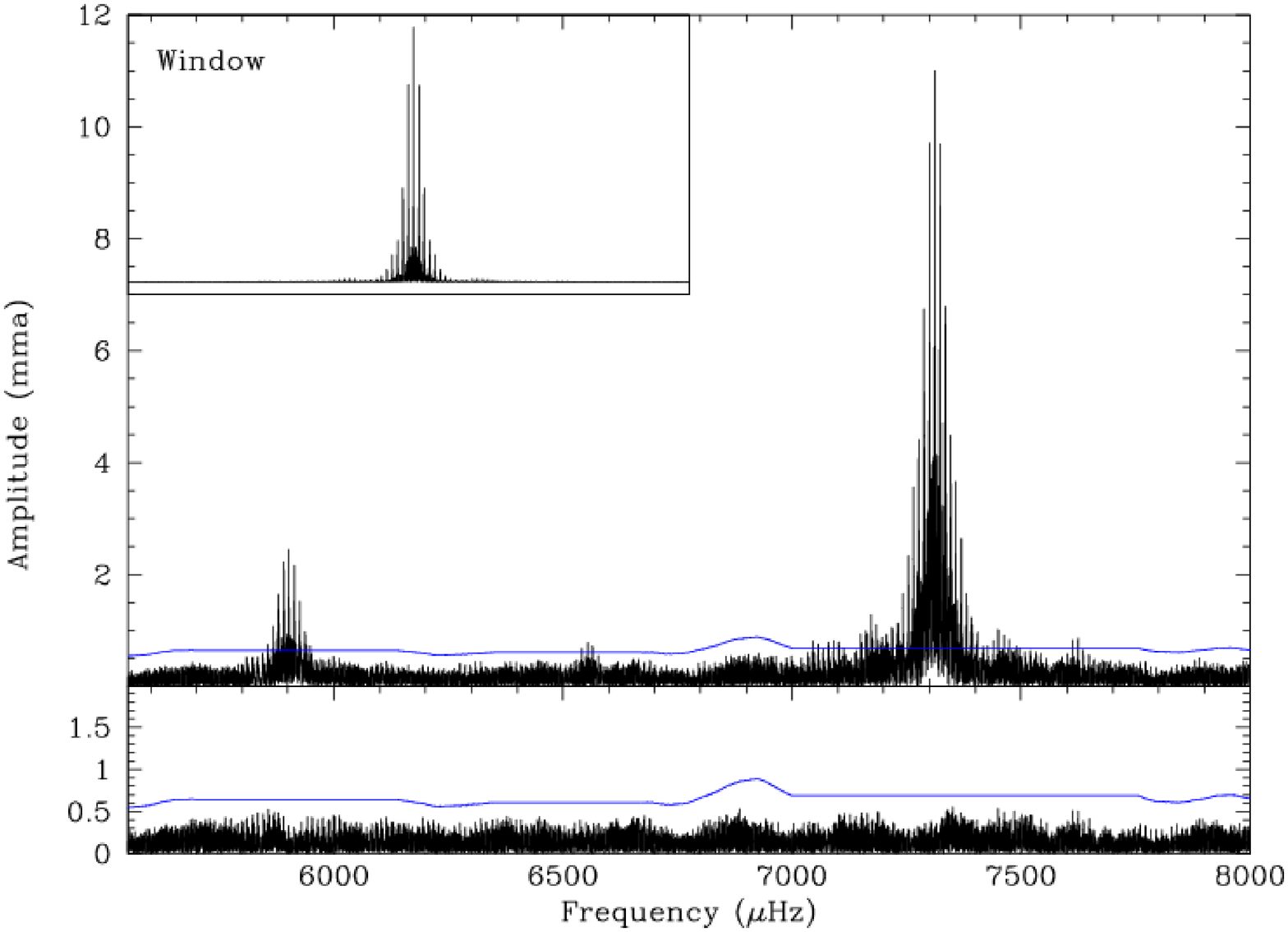,angle=-90,width=\textwidth}}
\caption{Temporal spectra for HS~0444. The top panel shows the
original FT, the bottom the residuals after prewhitening by three
frequencies, and the inset is the window function. The solid
(blue) line is the $4\sigma$ detection limit.\label{fig06}}
\end{figure}

\begin{figure}
 \centerline{\psfig{figure=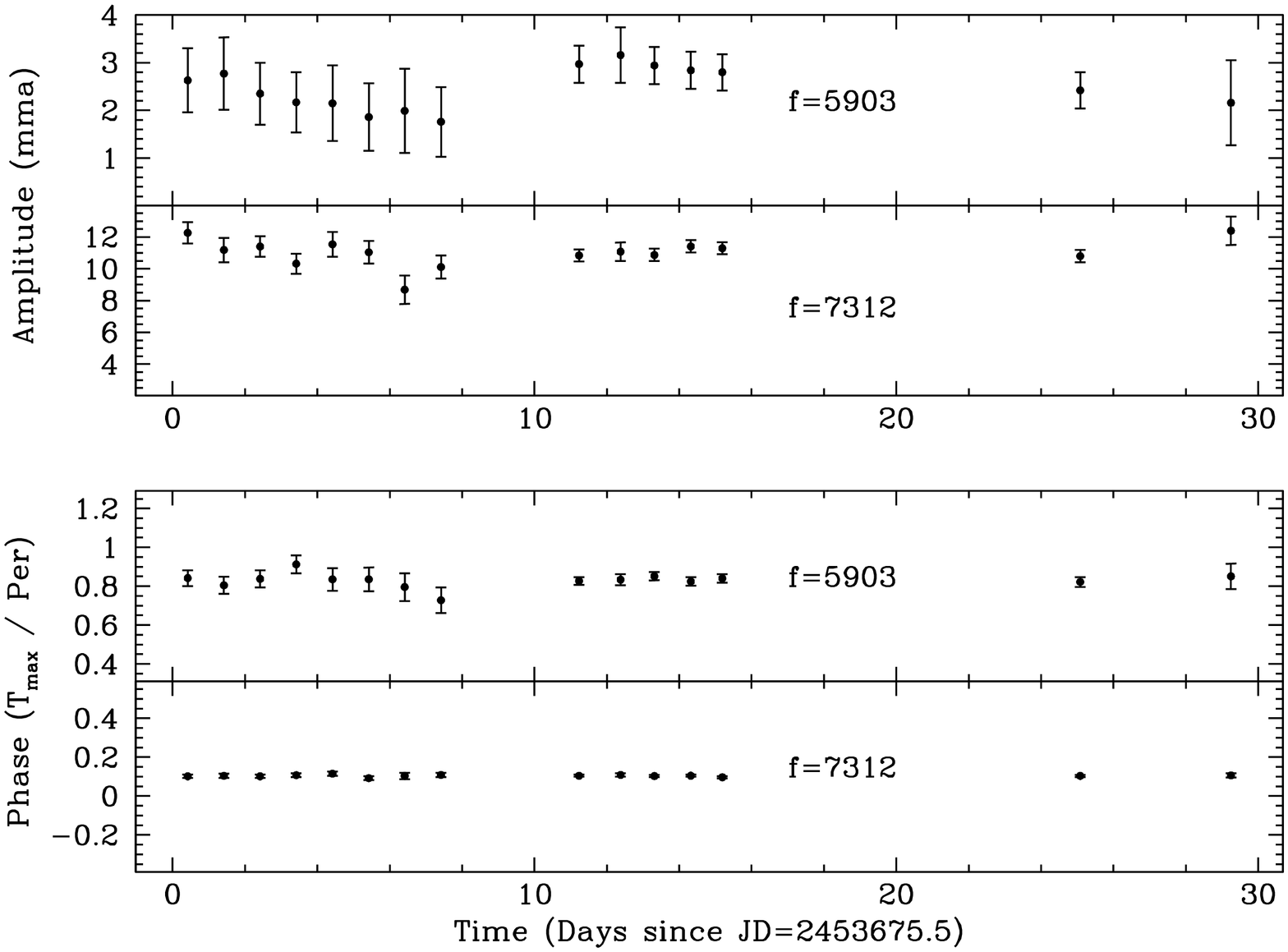,angle=-90,width=\textwidth}}
\caption{Amplitudes and phases for frequencies $f1$ and $f3$ of
HS~0444.\label{fig07}}
\end{figure}

{\bf HS~0444:} Early in our campaign, we concentrated on HS~0039,
leaving HS~0444 as a secondary target. Even from the relatively
short runs (Table \ref{tab02}), we could tell that HS~0444 was a
simple pulsator with just a few frequencies. In December, when
HS~0444 was better placed in the sky, we obtained longer
individual runs to ensure that we did not miss any low amplitude
peaks. The temporal spectrum of HS~0444 is shown in Fig.~\ref{fig06}. 
The top panel shows the original FT, the bottom panel
the residuals after prewhitening by three frequencies, and the
inset is the window function. The frequencies, periods and
amplitudes from our least-squares solution are provided in Table
\ref{tab04}. In Fig.~\ref{fig07}, we show the amplitudes and
phases of the two frequencies which were detectable each night.
The amplitudes and phases are relatively stable and do not show
any indications of additional unresolved frequencies.

\begin{table}
\caption{Periods, frequencies, and amplitudes for HS~0444. Formal
least-squares errors are in parentheses. }
\label{tab04}
\begin{tabular}{lccc}
\hline
ID & Frequency & Period  & Amplitude \\
& ($\mu$Hz) & (s) & (mma) \\ \hline
$f1$ & 5902.511 (0.012) & 169.41940 (0.00035) &   2.5 (0.2) \\
$f2$ & 6553.484 (0.037) & 152.59056 (0.00086) &   0.8 (0.2) \\
$f3$ & 7311.728 (0.003) & 136.76657 (0.00005) &  11.1 (0.2) \\ \hline
\end{tabular}
\end{table}

\section{Discussion}

\subsection{Comparison with the discovery data}

The goal of our observational program is to resolve the pulsation
frequencies for asteroseismic analysis. For the sake of comparison
with the discovery data, we calculate the temporal resolution as
$1/\Delta t$ with $\Delta t$ being the extent of the observations
in time \citep[see][]{kill99}. For the discovery data, we can
determine the temporal resolution from information provided in \O
01, and estimate the detection limit as twice the top of the noise
level in their FTs outside of the pulsations and window functions.

For HS~0039, our observations have a temporal resolution of
0.8~$\mu$Hz (excluding the December runs), which is $\approx
6.5\times$ better than the discovery data. The detection limit is
0.4~mma, which is about $5\times$ better than in \O 01. It is
difficult to determine whether the frequencies in HS~0039 have
changed since the discovery observations. \O 01's reported
frequencies (5.14, 5.48, and 4.27~mHz) roughly correspond to our
$f4$ or $f5$, $f7$, and $f2$, respectively, but \O 01's last
frequency of 5.21~mHz, detected in only a single observing run, is
not found in our data. As this frequency is wedged between
higher-amplitude frequencies, and \O 01's window function is
complex, it is difficult to judge the significance of its
detection, though it does look reasonable in their figures.
However, the main difficulty is the abundance of frequencies, many
of which have low amplitudes. The discovery data only detected
four frequencies, whereas we detect 14.  We estimate \O 01's noise
limit as $\approx 2$ mma;  if our limit were as high, we also
would have detected four frequencies.  It is therefore not
possible to ascertain the long-term stability of the pulsations
from just these two sets of data.

Additional data on HS~0039 were obtained using the ULTRACAM
multicolor instrument on the 4.2~m William Herschel Telescope in
2002 by Jeffery et al. (2004; hereafter J04). Two long ($\approx 8$~hr)
runs with very high signal-to-noise were obtained on consecutive
nights. We recover J04's eight frequency detections, six of which
are independent.  Two are aliases of $f7$ and one is an alias of
$f13$. J04 used multicolour photometry to estimate that $f8$ is an
$\ell =4$ mode. Though geometric cancellation should reduce
amplitudes for higher degree modes, the $(\ell,m) = (4,1)$ mode
can be reduced by as little as 30\%, depending on orientation
\citep{me1}. Since $f8$ is 18\% the amplitude of $f2$ (the
highest amplitude frequency), there is no problem in ascribing it
as an $\ell = 4$ mode. Additionally, as this region of the FT is
relatively uncrowded, the data obtained by J04 should have been
sufficient to make this determination. Although they do not claim
this identification with certainty, it seems reasonable and would
certainly be worth additional multicolour photometry or
time-series spectroscopy for confirmation.

For HS~0444, our observations have a temporal resolution of
0.4~$\mu$Hz, which is $\approx 14\times$ better than the discovery
data.  The detection limit is 0.6~mma, which is about $3\times$
better than in \O 01. Additionally, we have recovered the two
frequencies detected in the discovery data, to within the errors,
and uncover a single new frequency, at an amplitude below
their detection limit.

\subsection{Constraints on the pulsation modes}

In addition to improving the known pulsation spectra of these
stars, we wish to place observational constraints on the pulsation
modes.  The modes are mathematically described by spherical
harmonics with three quantum numbers, $n$ (or $k$), $\ell$, and
$m$.  Rotation can break the $m$ degeneracy by separating each
degree $\ell$ into a multiplet of $2\ell +1$ components, so
multiplet structure is a very useful tool for observationally
constraining pulsation degree \citep[see][]{wing91,me2,simon}.

For slow rotators, like most sdB stars are thought to be
\citep{heber99,heber00},
rotationally-split multiplets should be nearly equally spaced in
frequency. Such structure is, however, seldom observed in sdBV
stars and in neither case are multiplets detected in our
observations. Even though HS~0039 has 14 frequencies, the
frequency spacings are not regular.  Instead, the spacings are
distributed from 5 to 3325~$\mu$Hz, with no obvious groupings.  
For HS~0444, there are only three detected
frequencies, but the spacings are not similar, so there is no
multiplet structure in this star.

Another tool that can be used is the density of frequencies within
a given range. In resolved sdBV stars, we sometimes observe many
more pulsation modes than $\ell = 0$, 1, and 2 can provide,
independent of the number of inferred $m \neq 0$ frequencies.
Higher $\ell$ modes may be needed, but if so they must have a
larger amplitude than is measured because of the large degree of
geometric cancellation \citep{charp05a,me1}. A general guideline
would be one $n$ order per $\ell$ degree per 1000~$\mu$Hz
(\citet{char02} find an average spacing near 1440~$\mu$Hz), so the
temporal spectrum can accommodate three frequencies per
1000~$\mu$Hz without the necessity of invoking high-$\ell$ values
if no multiplet structure is observed. Filling all possible $m$ values,
the limit becomes nine frequencies per 1000~$\mu$Hz..

Obviously there is no need to invoke high-$\ell$ modes for
HS~0044.  Between 4900 and 5800~$\mu$Hz, HS~0039 has 9 of its 14
frequencies.  Therefore, HS~0039 has too large a frequency density
to exclude $\ell\geq 3$ modes, particularly with the absence of
any obvious multiplets. This supports the identification of an
$\ell = 4$ mode by J04.

\section{Group Properties}

\subsection{Data sources}

Since the discovery of the EC~14026 class of pulsating sdB stars
in 1997 \citep{kill97}, there have been three areas of emphasis
for observations: 1) to discover more pulsators; 2) to resolve the
pulsations using long time-base campaigns, sometimes at multiple
sites; and 3) to obtain high signal-to-noise observations over
short time intervals.  In recent years, multicolour photometry and
time-series spectroscopy have been obtained as additional tools 
for mode identification.

For this paper, we will concentrate on the second point above, and
examine pulsators for which a considerable effort has been
expended to resolve the pulsation frequencies.  
We do this because it has been our area of emphasis,
we feel that it is an important component in applying asteroseismology
to sdB stars, and most importantly, we have data
for all the stars except those from Kilkenny et al. (2002, 2006a, 2006b).
Though this will
not be a complete sample of sdBV follow-up observations, 
we can perform uniform tests upon them (except as noted above) for
intercomparison.
%This has been our
%area of emphasis, and we have observed a large number of known
%sdBV stars except those recently identified by
%\citet{kill06a,kill06b}.

Table~\ref{tab05} provides a list of studies that have thoroughly
investigated the pulsations of sdBV stars. Some stars, such as
Feige~48, PG~0014, PG~1219, and PG~1605, have received extensive
observations over the course of many years, while most have only
been observed during a single campaign. Column 1 of that table
lists the full name of each pulsator.  Columns 2 and 3 display the
dates of observations and the number of hours observed.  The final
two columns display the observing sites and references for each
star.  The references in Table~\ref{tab05} are the basis for our
analysis below, but are not necessarily a complete record of the
detailed observations of each star.

\begin{table}
\caption{List of follow-up observations of pulsating sdB stars. A
dagger$^{\dagger}$ indicates observations that we were involved in and NA
indicates information that was not available. Observing sites: 1)
Suhora 0.6~m; 2) Baker 0.4~m; 3)  CTIO 1.5~m; 4) SAAO 1.9~m; 5)
Fick 0.6~m; 6) MDM 1.3~m; 7) McDonald 2.1~m; 8) McDonald 0.9~m; 9)
MDM 2.4~m; 10) Whole Earth Telescope Campaign, 11) Multisite
campaign by Reed et al., 12) Other campaign. Column 5 references
are: a) Baran et al. 2005; b) Reed et al. 2006b; c) Kilkenny et
al. 2006a; d) Kilkenny et al. 2006b; e) Reed et al. 2004; f) this
work; g) Reed et al. 2006a; h) Silvotti et al. 2002a; i) Reed et al.
2006c; j) Zhou et al. 2006; k) O'Donoghue et al. 1998a; l)
Vu\u{c}kovi\'{c} et al. 2006; m) Reed et al. 2007; n) Kilkenny et
al. 2002; o) Harms, Reed, O'Toole 2006; p) Silvotti et al. 2006;
q) Kilkenny et al. 2003; r) Kilkenny et al. 1999; U) unpublished.\label{tab05}}
\begin{tabular}{lcccc} \hline
Target & Inclusive Dates & Hours Observed &  Sites & References  \\ \hline
Balloon090100001 & 17 Aug. - 19 Sep. 2004 & 125 & 1  & a\\
$^{\dagger}$ & 8 Aug. - 30 Sep. 2005 & NA & 2, 12 & U\\
EC~05217-3914$^{\dagger}$ & 6 - 15 Nov. 1999 & 59 &  3, 4 & b \\
EC~14026-2647 & July 2003 & NA & 4 & c\\
EC~20338-1925 & 23 Jul. - 26 Sep., 1998 & 45.9 & 4 & d \\
        & June 2004 (2 nights) & 12 & 4 & c\\
Feige~48$^{\dagger}$ & 1998 -- 2006 & $>500$ &  2, 5, 6, 7, 8, 10, 11 & e \\
HS~0039+4302$^{\dagger}$ & 15 Nov. - 14 Dec.  2005 & 91 & 6 & f \\
HS~0444+0458$^{\dagger}$ & 15 Nov. - 14 Dec. 2005 & 63 & 6 & f \\
HS~1824+5745$^{\dagger}$ & 25 May  - 11 Jul.  2005 & 127 & 6, 9  & g \\
HS~2149+0847 & July 2003 (4 nights) & NA & 4 & d\\
             & June 2004 (10 nights) & NA & 4 & d\\
HS~2151+0857$^{\dagger}$ & 18 Jun.  - 11 Jul. 2005 & 42 & 6, 9 & g \\
HS~2201+2610$^{\dagger}$ & 17 Sep.  - 4 Oct. 2000 & 95.0 &  5, 12 & h \\
KPD~1930+2752$^{\dagger}$ & 11 - 16 Jul. 2002 & 38 & 7, 11 & i\\
       $^{\dagger}$       & 15 Aug.  - 9 Sep.  2003 & 246.5 &  10 & U\\
KPD~2109+4401$^{\dagger}$ & 12 Sep.  - 14 Oct. 2004 & 182.6 &  2, 6, 11 & j \\
PB~8783 & 8 - 22 Oct. 1996 & 183 & 12 & k \\
PG~0014+182$^{\dagger}$ &  8 - 20 Oct. 2004 & 142 &  6, 10 & l \\
PG~0048+091$^{\dagger}$ & 26 Sep. - 11 Oct., 2005 & 167 & 9, 11 & m \\
PG~0154+182$^{\dagger}$ & 6 - 14 Oct. 2004 & 28.4 &  6 & g \\
PG~1047+003 & 17 Feb. - 2 Mar. 1998 & 98 & 12 & n \\
PG~1219+534$^{\dagger}$ & 2003 -- 2006 & $>200$ &  2, 6, 7 & o \\
PG~1325+101$^{\dagger}$ & 3 Mar.  - 3 Apr. 2003 & 264 &  2, 12 & p \\
PG~1336-018$^{\dagger}$ & 3 - 20 April, 1999 & 172 &  10 & q \\
          $^{\dagger}$  & 14 Apr.  - 1 May 2001 & 288 &  10 & U\\
PG~1605+072$^{\dagger}$ & 1997 -- 2002 & $>400$ & 4, 5, 10, 11 & r \\
PG~1618+563$^{\dagger}$ & 17 Mar. - 1 May  2005 & 200.5 &  2, 7, 9, 11 &
m\\ \hline
\end{tabular}
\end{table}

\subsection{The pulsation content}

In Table \ref{tab06}, we assemble data on the various sdBV stars.
Column 1 lists an abbreviated name, which we will use hereafter.
Columns 2 -- 4 give the total number of detected frequencies, and
the numbers with high and low amplitudes (these and other
quantities in the table are discussed more thoroughly below). In
columns 5 and 6, we show the temporal resolution $1 / {\Delta t}$
and the noise limit in mma as inferred from the analysis in each
paper (\S 4.1). Column 7 displays our judgment about whether the
frequencies were completely resolved. Of the 23 stars in Table~\ref{tab07}, 
18 are likely resolved. The other five either have too little data (EC~20338 and
PB~8783), or have (some) pulsation amplitudes that are variable on
time-scales too short for frequency resolution (BA09, KPD~1930, and
PG~0048). The effective temperatures
and gravities are listed in Columns 8 and 9;  quantities in
parentheses are estimates (described in \S 5.3).  
The next two columns display the total
power in the resolved frequencies and the largest detected
amplitude. The last column provides references for the spectroscopically
determined values of $\log g$ and $T_{\rm eff}$.

\begin{figure}
 \centerline{\psfig{figure=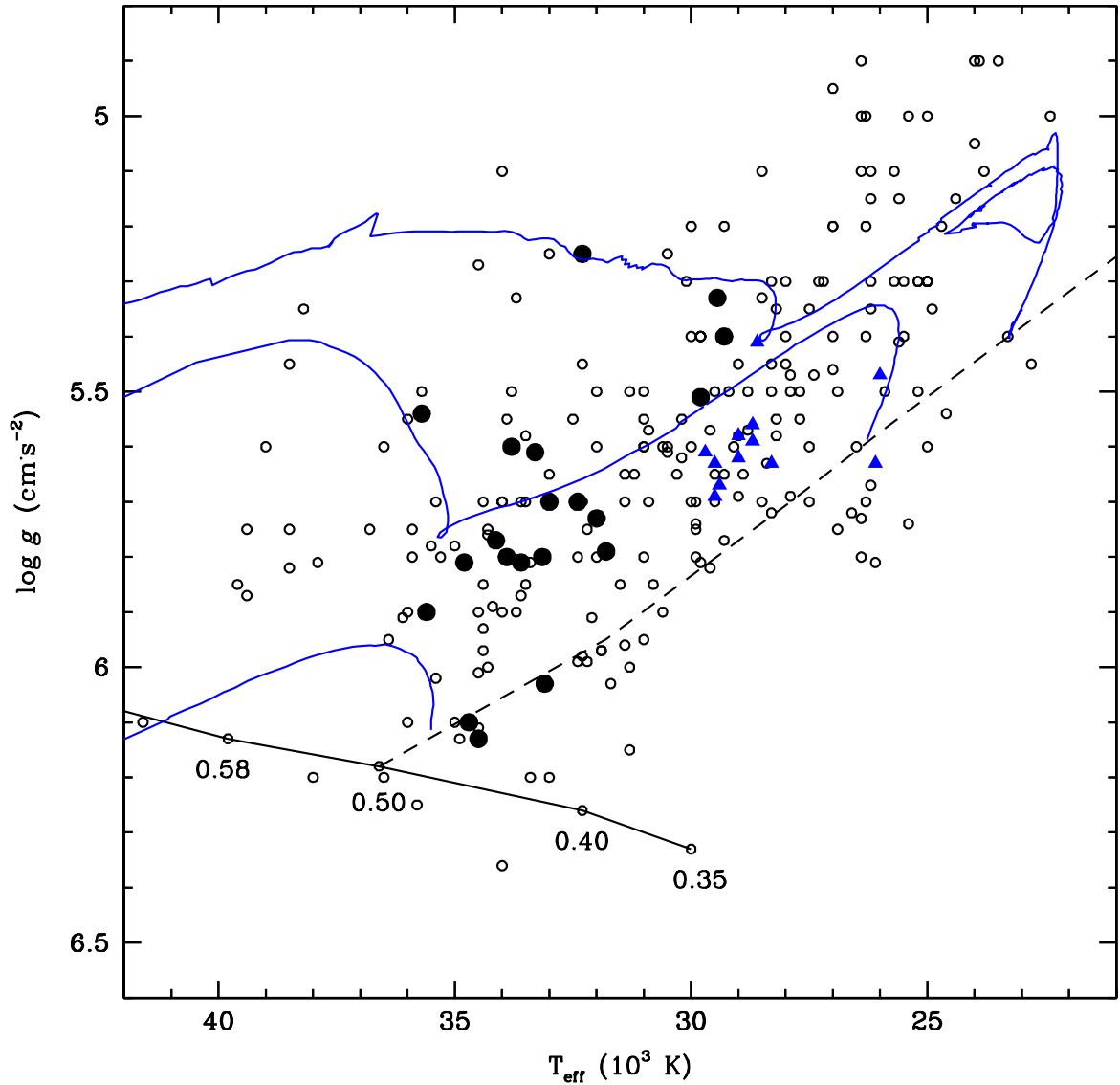,width=\textwidth}}
\caption{A $\log g - T_{\rm eff}$ diagram showing the location of
sdB stars. Filled circles are EC~14026-type pulsators, filled
(blue) triangles are PG~1716-type pulsators, and open circles are
non-pulsating sdB stars. The dashed line is the zero-age extended
horizontal branch, the black line is the zero-age helium main
sequence with open circles indicating the total mass. The
evolutionary tracks (solid blue lines) were produced using ISUEVOS
\citep{me2}.\label{fig08}}
\end{figure}

Figure \ref{fig08} displays $\log g$ against $T_{\rm eff}$ for the
20 stars in Table~\ref{tab06} with previously determined values.
These are shown as filled circles.  In addition, PG~1716-type
pulsators are shown as filled (blue) triangles, and non-pulsators
are shown as open circles. The solid (black) line is the zero-age
helium main sequence with masses marked, the dashed line is the
zero-age extended horizontal branch, and evolutionary tracks
\citep[from][]{me2} are shown as solid (blue) lines. The coolest
track has a hydrogen envelope thick enough for shell fusion, while
the hotter two do not.

\begin{figure*}
 \centerline{\psfig{figure=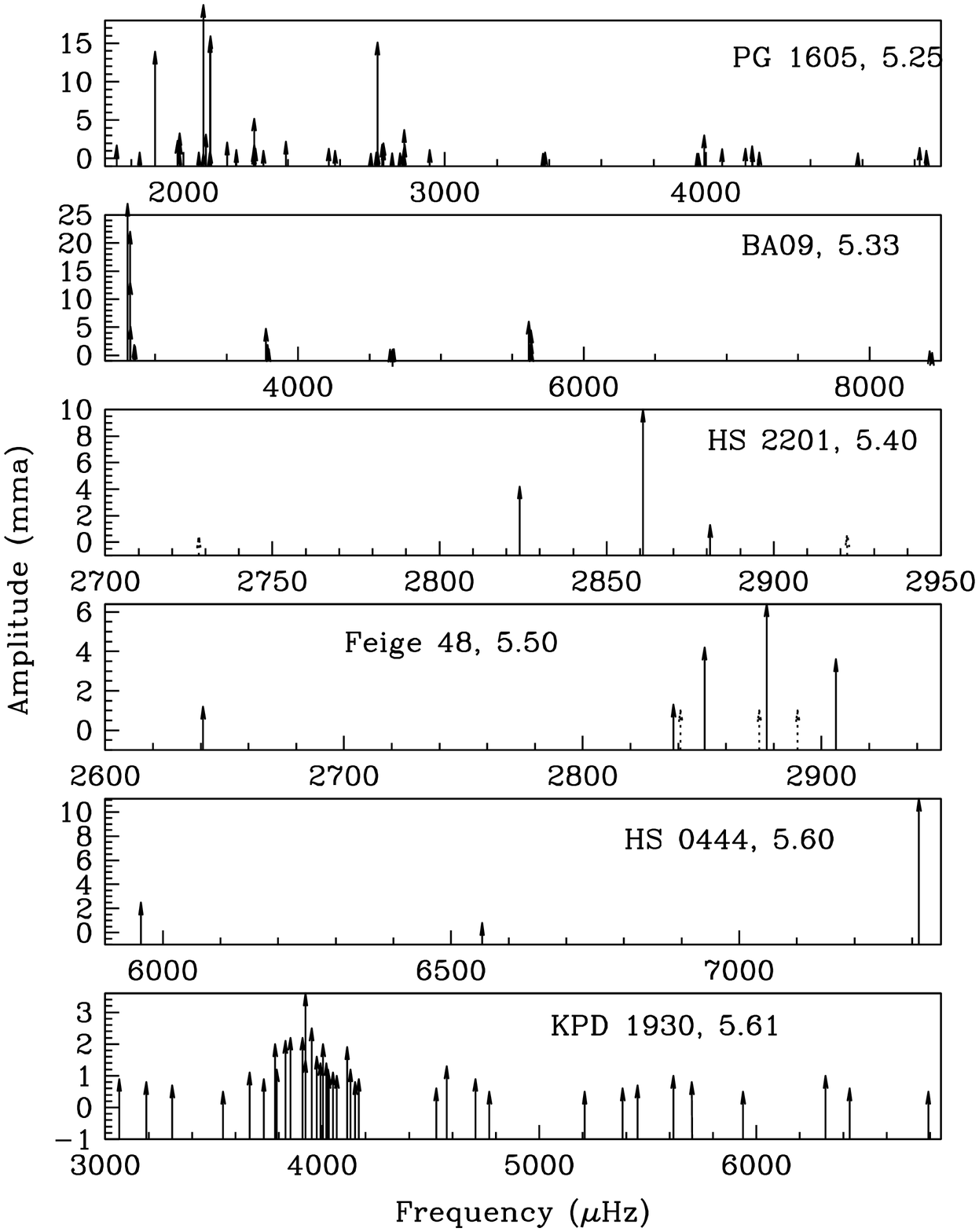,width=3.7in}\psfig{figure=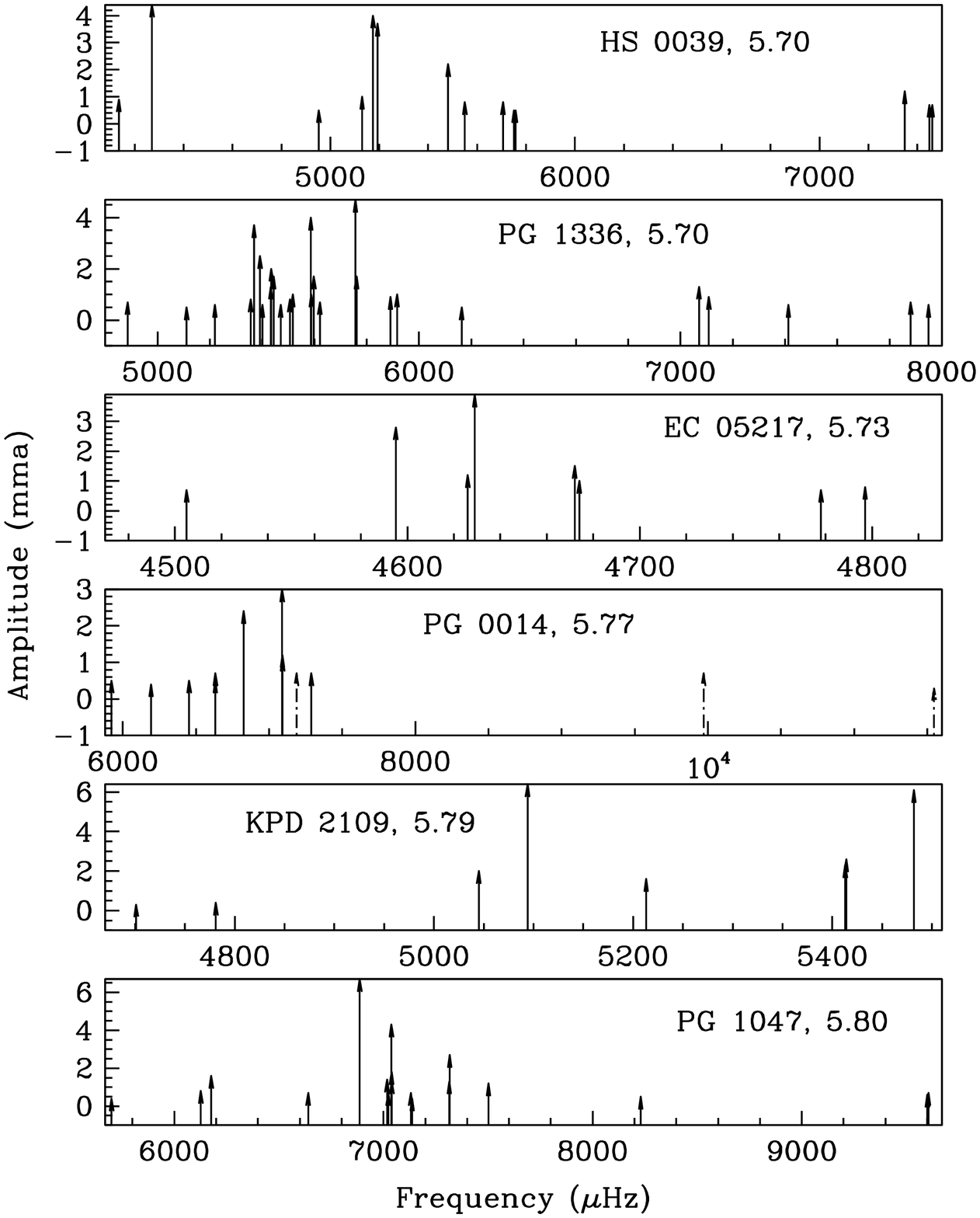,width=3.7in}}
% \centerline{\psfig{figure=schem1a.ps,width=3.5in}\psfig{figure=schem1b.ps,width=3.5in}} COLOR FIGURES
\caption{Schematic representation of the temporal spectra of sdBV
stars with follow-up data ordered by $\log g$ (from lowest to
highest). Blue arrows indicate frequencies that are not regularly
detected and green arrows for PG~0014 indicate frequencies only
detecting using ULTRACAM.  Shown here are the stars with lower
gravities.\label{fig09}}
\end{figure*}

\begin{figure*}
 \centerline{\psfig{figure=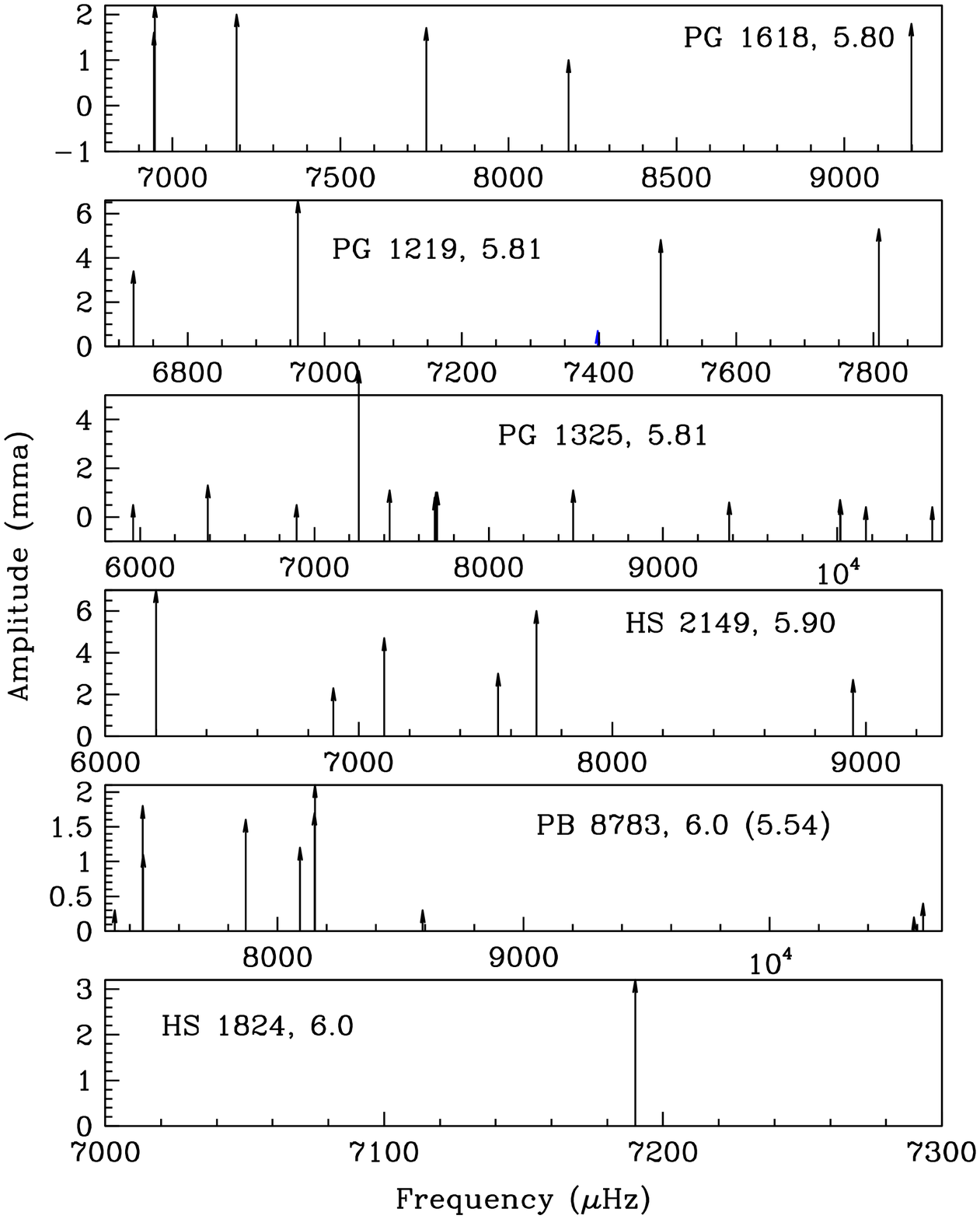,width=3.7in}\psfig{figure=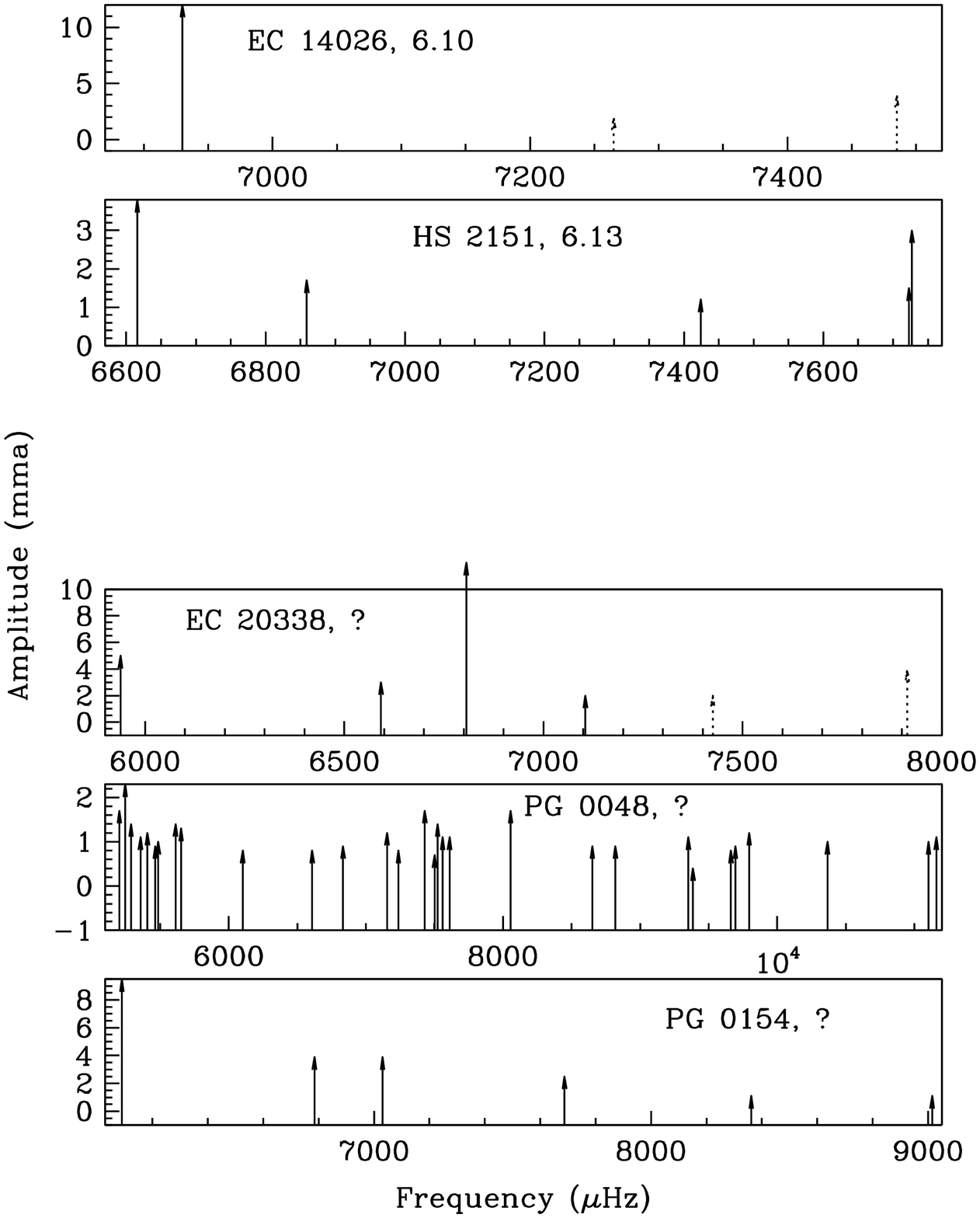,width=3.7in}}
% \centerline{\psfig{figure=schem1c.ps,width=3.5in}\psfig{figure=schem1d.ps,width=3.5in}} COLOR FIGURES
\caption{Same as Fig.~\ref{fig09}, but for stars with higher
gravities.  The last three panels are for the stars without a
previously determined value of $\log g$.\label{fig10}}
\end{figure*}

Figures~\ref{fig09} and \ref{fig10} show schematic representations
of the temporal spectra of all 23 stars, ordered by $\log g$, with
the three stars without spectroscopically constrained values at the end. 
The dotted
arrows (blue in the electronic version) indicate frequencies which
are only observed occasionally. PG~0014 has dashed arrows (green
in the electronic version), indicating frequencies that were only
observed using ULTRACAM. To make
low-amplitude frequencies visible in the plots, the  vertical axes
may begin below zero and
are scaled so that the highest amplitude peak touches the top
line, except for PG~1605, BA09, PG~1325, and EC~20338.  All of the
latter have one high-amplitude peak that would make the others too
small on the plot. Those frequencies are indicated with arrows
which pass beyond the top of the plot, though still not to scale.
Note that all panels are plotted at different scales.  BA09 shows
a mixture of short- (EC~14026-type), and long-period
(PG~1716-type) oscillations;  only the former are shown.

\begin{table}
\caption{Pulsation properties of EC~14026-type pulsators for which
follow-up data has been obtained. References for the spectroscopic
measurements are as follows: 1) Oreiro et al. 2004; 2) Koen et al.
1999a; 3) O'Donoghue et al. 1997; 4) Koen et al. 1998b; 5) \O
stensen et al. 2001a; 6) \O stensen et al. 2000b; 7) Bill\' eres
et al. 2000; 8) Koen 1998; 9) Brassard et al. 2001; 10) O'Donoghue
et al. 1998b; 11) Koen et al. 1999b; 12) Silvotti et al. 2002b; 13)
Kilkenney et al. 1998; 14) Koen et al. 1998a; 15) Silvotti et al.
2000; U) Unpublished. Parameters in parentheses are inferred using color and
pulsation frequencies. NA indicates information that is not
available. BA09 also has longer period (PG~1716-type) pulsations and
combination frequencies
which are not included in this table.\label{tab06}}
\begin{tabular}{lccccccccccc} \hline
Star & Total & High & Low & $1/\Delta T$ & Limit & Resolved? & $T_{\rm eff}$ & $\log g$ &
Power & A$_{\rm max}$ & Refs \\
     &  (\#)  & (\#)  & (\#) & ($\mu$Hz) & (mma) &  & (K) & ($cm\cdot s^{-2}$)& (mma$^2$) & (mma) & \\ \hline
BA09 & 19 & 3 & 16 & 0.4 & 0.5 & No & 29446 & 5.33 & 4103.92 & 57.7  & 1 \\
EC~05217 & 8 & 6 &  2 & 0.9 & 1.3 & No & 32000 & 5.73 & 29.36 & 3.9 & 2 \\
EC~14026 & 3 & 1 & 2 & NA & NA & Yes & 34700 & 6.10 & 164 & 12 & 3 \\
EC~20338 & 5 & 3 & 2 & 1.2 & 0.8 & No & (35500) & (5.8) & 774.43 & 26.6 & U\\
Feige~48 & 8 & 3 & 5 & 0.8 & 0.1 & Yes & 29500 & 5.50 & 69.73 & 6.4 & 4\\
HS~0039 & 14 & 6 & 8 & 0.8 & 0.4 & Yes & 32400 & 5.70 & 60.15 & 4.4 & 5 \\
HS~0444 & 3 & 2 & 1 & 0.4 & 0.6 & Yes & 33800 & 5.60 & 130.1 & 4.4 & 5 \\
HS~1824 & 1 & 1 &0& 0.25 & 0.48 & Yes & 33100 & 6.03 & 11.56 & 3.4 & 5 \\
HS~2149 & 6 & 6 & 6 & NA & NA & Yes & 35600 & 5.90 & 28.67 & 7.0 & 6 \\
HS~2151 & 5 & 5 & 0 &0.5 & 0.53 & Yes & 34500 & 6.13 & 30.02 & 3.8 & 5 \\
HS~2201 & 5 & 2 & 3 & 0.01 & 0.5 & Yes & 29300 & 5.40 & 119.94 & 10.8  & 6 \\
KPD~1930 & 39 & 31 & 8 & 0.5 & 0.8 & No & 33300 & 5.61 & 74.54 & 3.6 & 7 \\
KPD~2109 & 8 & 6 & 2 & 0.4 & 0.29 & Yes & 31800 & 5.79 & 97.0 & 6.4 & 8 \\
PB~8783 & 10 &  6 &  4 & 0.8 & 0.1 & No & 35700 & 5.54 & 16.13 & 2.1 & 3 \\
PG~0014 & 13 & 8 & 5 & 0.9 & 0.48 & Yes & 34130 & 5.77 & 20.37 & 3.9 & 9 \\
PG~0048 & 30 & 29 & 1 & 0.28 & 0.8 & Yes & (34000) & (5.75) & 42.16 & 2.3 & U\\
PG~0154 & 6 & 4 & 2 & 1.4 & 0.76 & Yes & (35000) & (5.8) &  129.34 & 9.5 & U\\
PG~1047 & 18 & 6 & 12 & 0.8 & 0.2 & Yes & 33150 & 5.80 & 87.0 & 6.7 & 10 \\
PG~1219 & 6 & 4 & 2 & 0.8 & 0.6 & Yes & 33600 & 5.81 & 108.18 & 6.6 & 11 \\
PG~1325 & 14 &  6 &  8  & 0.5 & 0.8 & Yes & 34800 & 5.81 & 744.27 & 27.1 & 12 \\
PG~1336 & 27 & 13 & 14 & 1.0 & 0.25 & Yes & 33000 & 5.70 & 83.75 & 4.7 & 13 \\
PG~1605 & 55 & 5 & 50 & 0.8 & 0.5 & Yes & 32300 & 5.25 & 1802.1 & 27.4 & 14 \\
PG~1618 & 6 & 6 & 0 & 0.3 & 0.59 & Yes & 33900 & 5.80 & 18.53 & 2.2 & 15 \\ \hline
\end{tabular}
\end{table}

These two figures show the enormous variety of amplitudes and
frequencies detected in sdBV stars. There are only four
high-amplitude (here $A > 20$~mma) pulsators known and they have a
great range (for sdB stars) of gravities and temperatures. There
are pulsators with 20+ frequencies that have similar temperatures
and gravities to stars with 5 frequencies (e.g., BA09 \& HS~2201).
If one looks at only the range of gravities from $\log g = 5.6$ to
5.7 (about $1\sigma$ in error), there are two stars with more than
20 frequencies yet one star with only three frequencies (but
higher amplitudes!).

\subsection{Observational tests and trends}

In an attempt to bring order to the class as a whole and to
find trends in the observational properties, we have organized
the data in several ways which have benefited studies of other
variable stars. In this subsection we will show the results along
with some motivation, but leave in-depth interpretations to the next
subsection.

\begin{figure*}
 \centerline{\psfig{figure=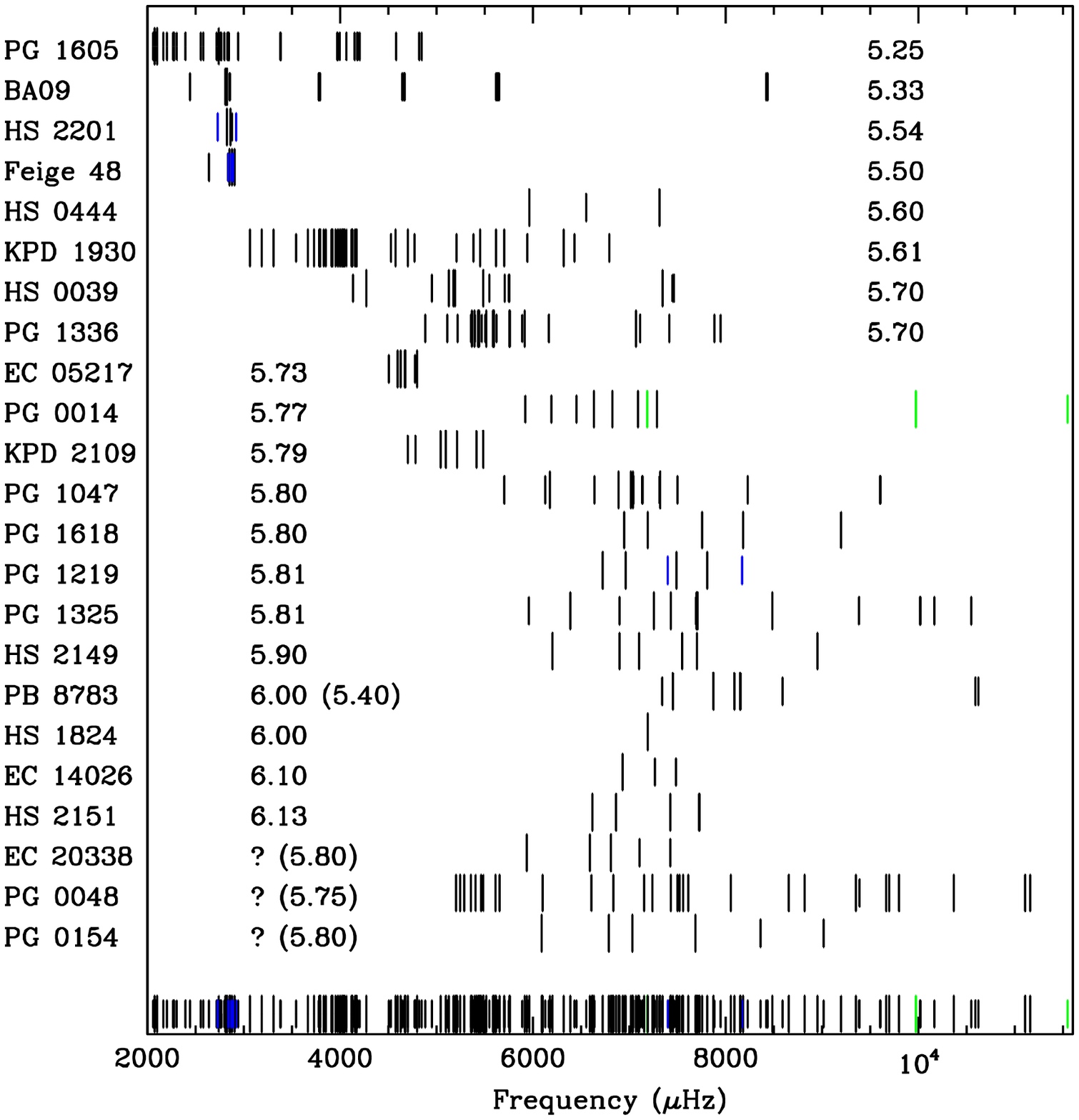,width=3.7in}\psfig{figure=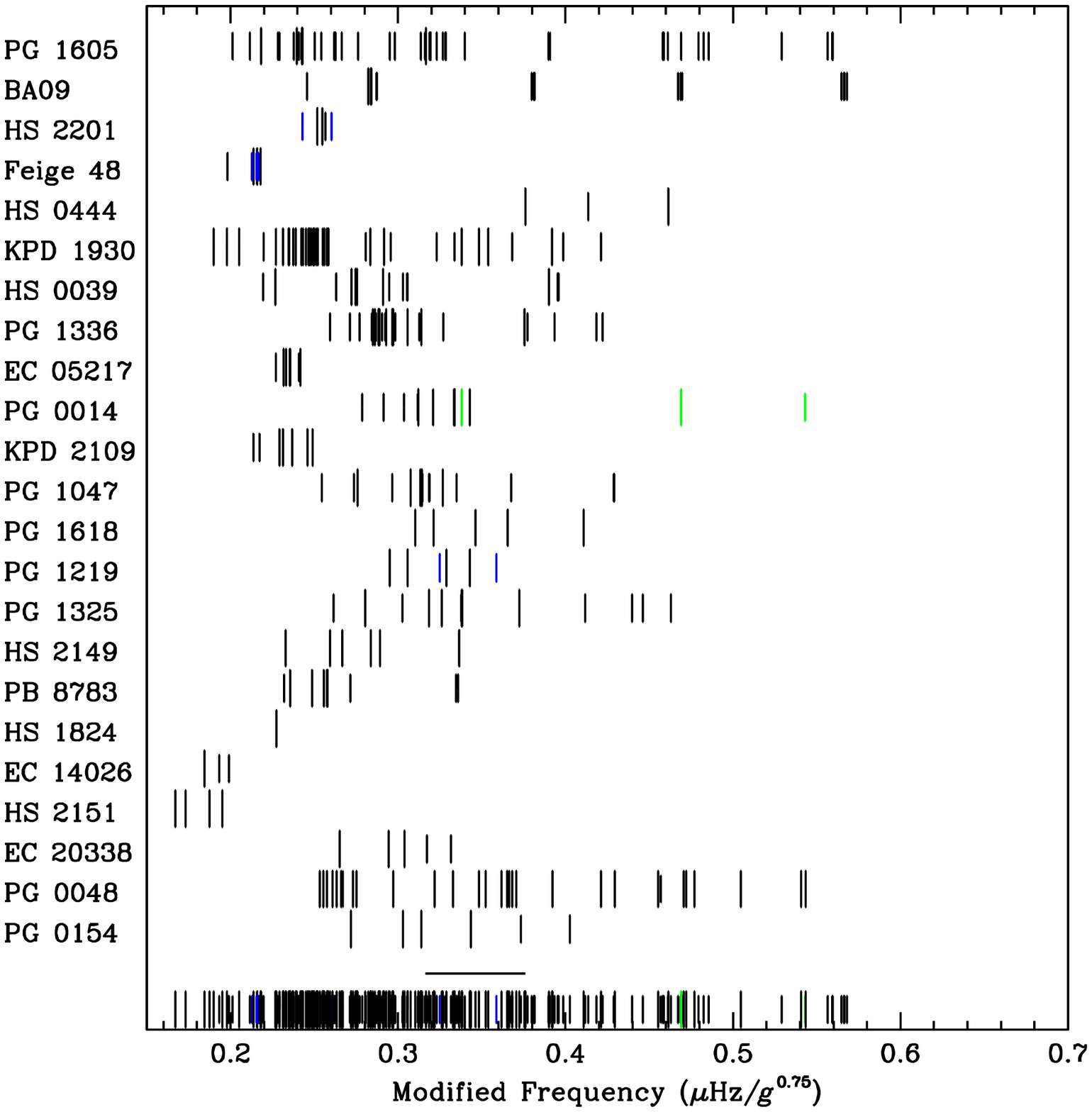,width=3.7in}}
\caption{Schematic representations of pulsation frequencies
organized by $\log g$. The lines used to indicate frequencies have
two lengths, with the shorter lengths indicating frequencies with
amplitudes a factor of 5 smaller than the highest amplitude. The
right panel has a frequency correction for $\log g$ and both
panels include a summation of all frequencies at the
bottom.\label{fig11}}
\end{figure*}

{\bf Frequency groupings:}
Our first arrangement was to put the frequencies shown in Figs.~\ref{fig09}
and \ref{fig10} onto a common frequency scale and make a correction
for gravity. Because $p-$mode periods are inversely proportional to the square
root of the density, we expect that stars with lower $\log
g$ have longer periods (shorter frequencies).  This is largely
observed in the left panel of Fig.~\ref{fig11}. In the right
panel, we have rescaled pulsation frequencies by $1/g^{0.75}$
(assuming constant mass)
which is an adjustment for both size and density using just $\log g$. For
the three sdBV stars without measured gravities, we estimated the
gravity from the position of the shortest frequency compared to
pulsators with known gravities. These assumed gravities are
provided in parentheses in Fig.~\ref{fig11} and Table
\ref{tab06}. Such a study of pulsating white dwarfs revealed
groups of frequencies which could then be related to individual
modes \citep{clem94}. However, as evidenced by the summation of
the right panel, no such groupings occur. The horizontal line just above
the summation frequencies (right panel of Fig. \ref{fig11}) shows
the effect of an error of $\log g = 0.05$ in the rescaling. It is
therefore possible that any groups are being smeared out by
measurement errors in $\log g$. We attempted to correct for this by
fixing the lowest modified frequency to a given value, but no
corrections or fixed reference values of $\log g$ show reasonably-separated 
grouping that could be of use.

{\bf Relative pulsation amplitudes:}
The line lengths in Figs.~\ref{fig09} through \ref{fig11}
indicate another feature that is observed about half the time; one
or two amplitudes are significantly higher than the rest.  To
parameterize this phenomenon, we have denoted frequencies whose
amplitudes are within a factor of five of the highest amplitude as
``high'' amplitudes;  the remainder are ``low'' amplitudes.  The
numbers that fall into each category, along with the highest
amplitude observed ($A_{\rm max}$) for each star, are provided in
Table~\ref{tab06}. The factor of five was chosen so that all
amplitudes greater than 10~mma for PG~1605 and BA09 would fall
into the high category. Additionally, we use the published
(typically average) amplitudes, even though some vary by large
amounts (discussed below).

\begin{figure*}
\centerline{\psfig{figure=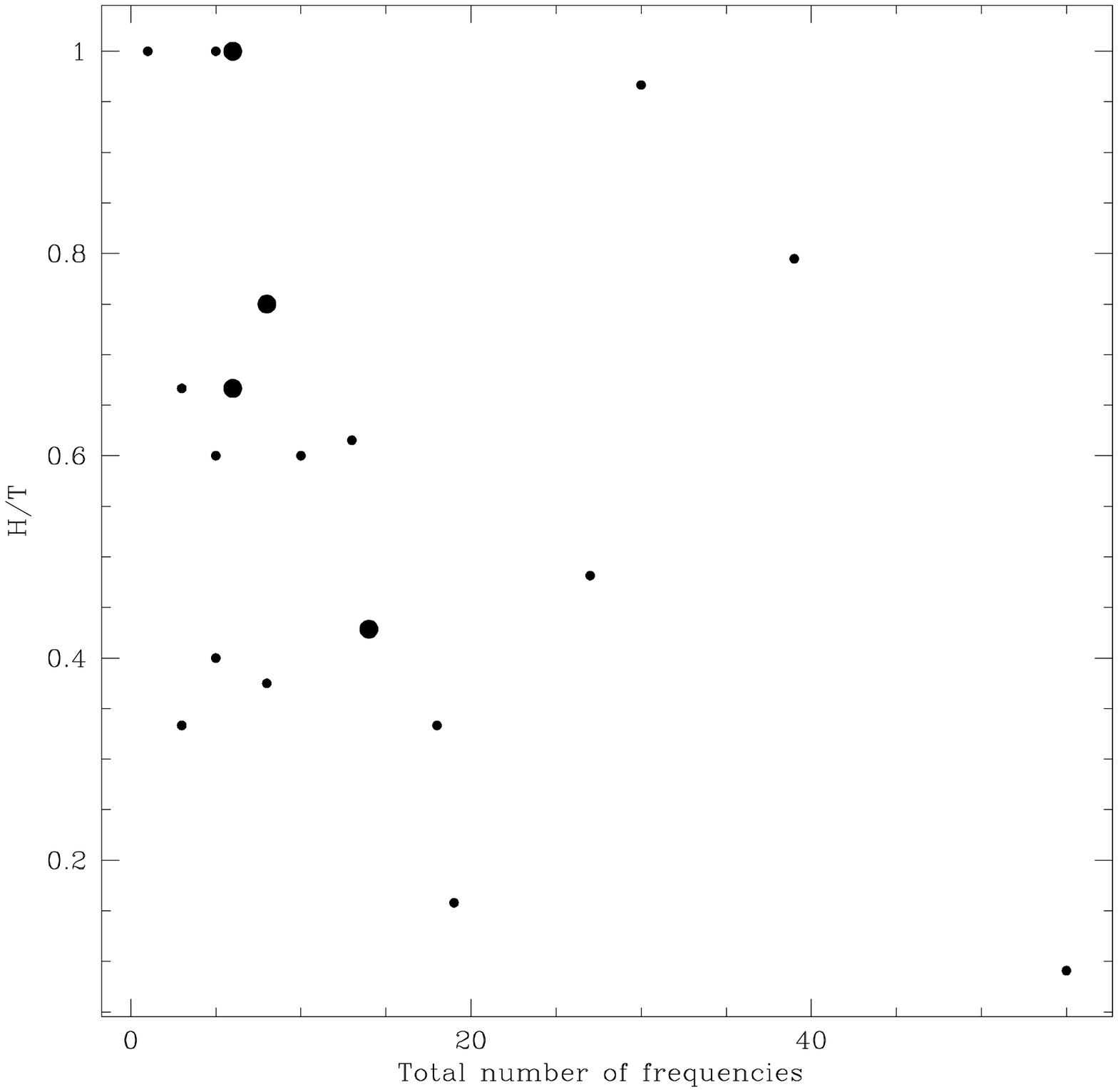,width=3.7in}\psfig{figure=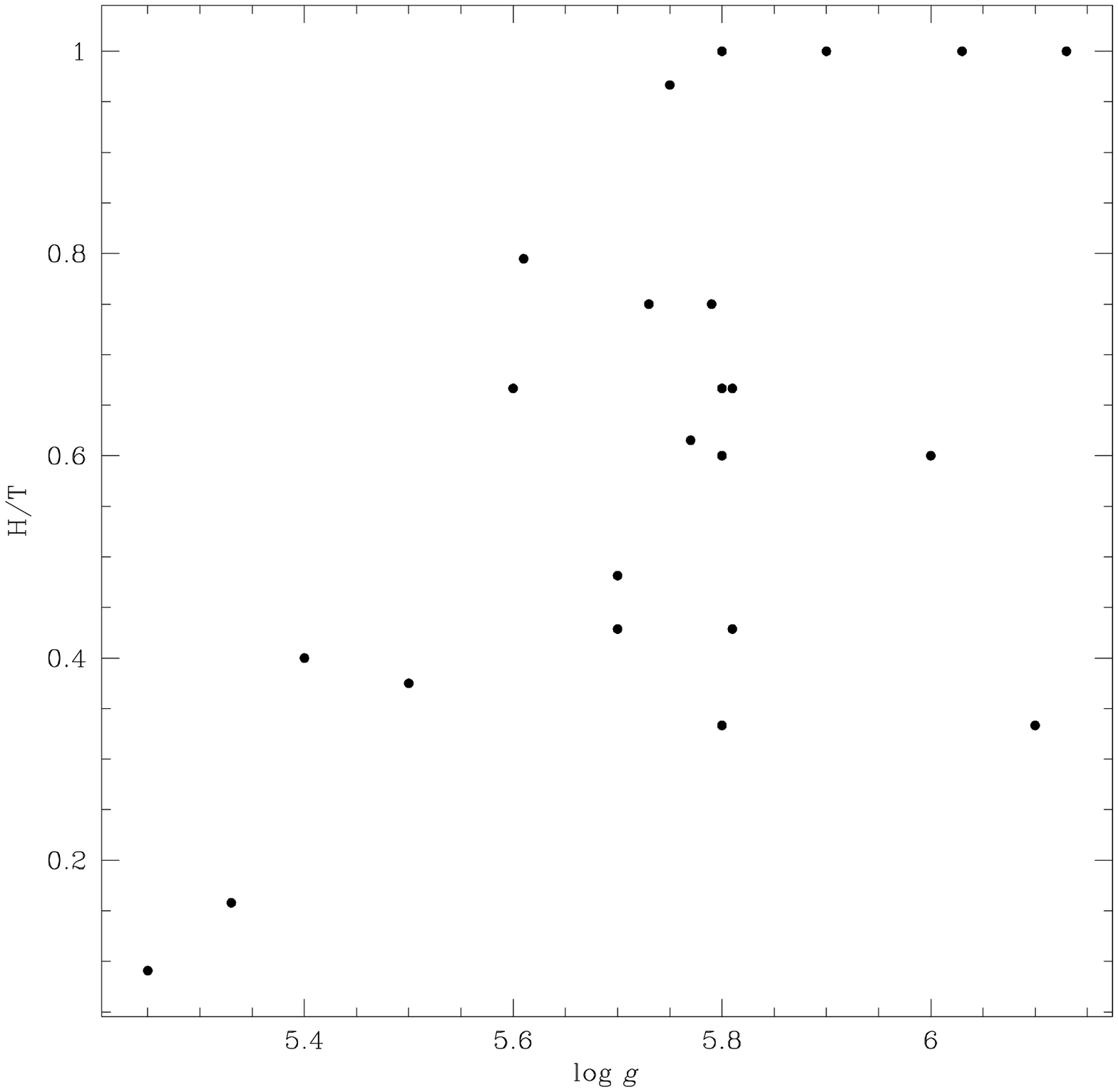,width=3.7in}}
\caption{Comparison of number of ``high'' amplitude frequencies
(H) to the total number of frequencies (T). Double sized points
indicate two stars with the same quantities. The right panel
compares this ratio with the surface gravity.\label{fig12}}
\end{figure*}

The left panel of Fig.~\ref{fig12} shows the ratio of high (H)
to total (T) number of frequencies against the total number of
frequencies;  the values of which are also provided in Table \ref{tab06}. One
might suspect that if a star has one high amplitude, all
amplitudes are relatively higher and thus easier to detect. This
appears not to be the case, since the points make a scatter
diagram.  In the right panel of Fig.~\ref{fig12}, we show the
H/T ratio against $\log g$.  It might be expected that gravity
plays a factor, in that it is easier to pulsate radially at lower
gravities.  The values of H/T are indeed positively correlated
with gravity, but the correlation coefficient is only 0.59.

\begin{figure*}
 \centerline{\psfig{figure=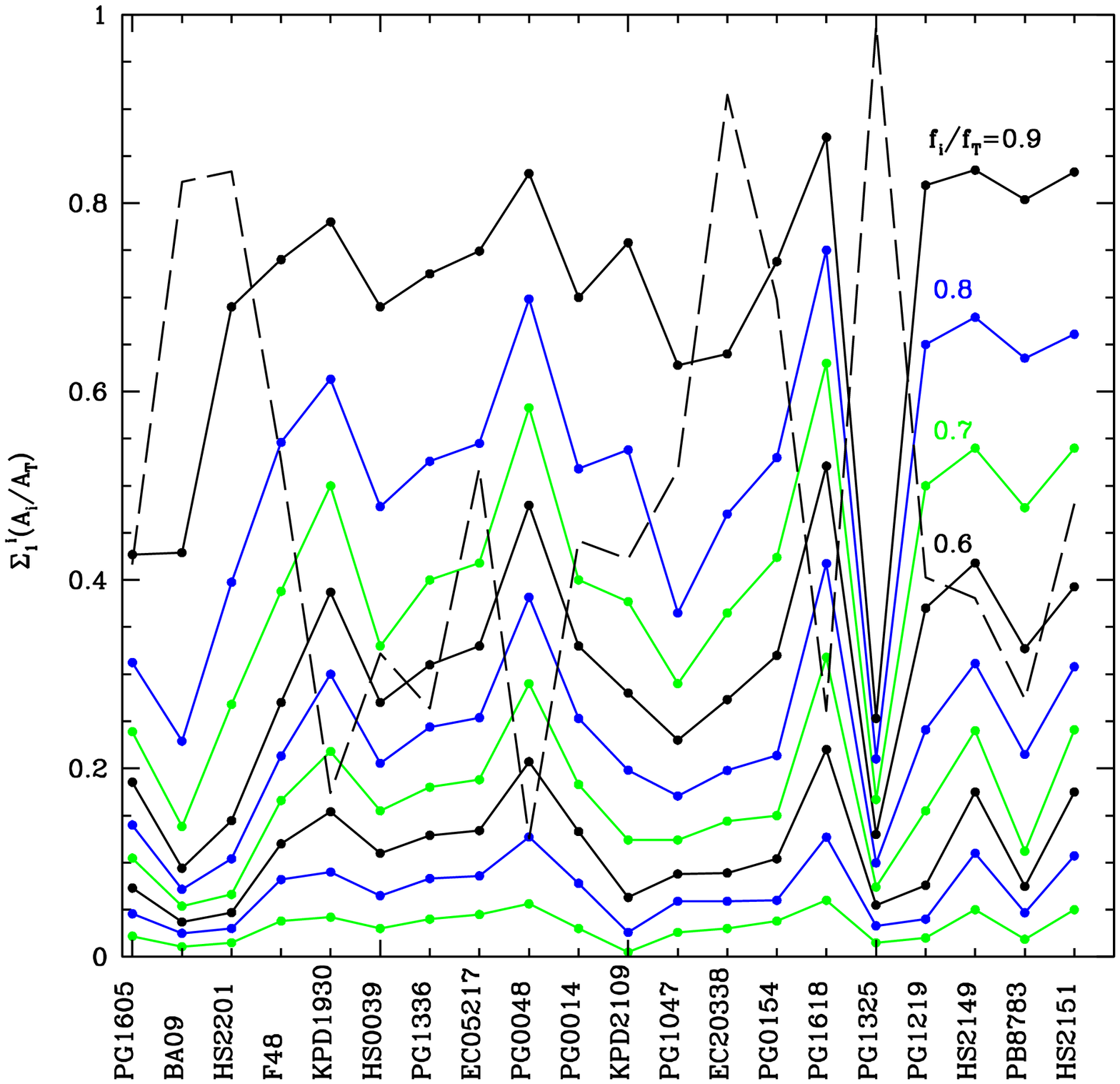,width=3.7in}\psfig{figure=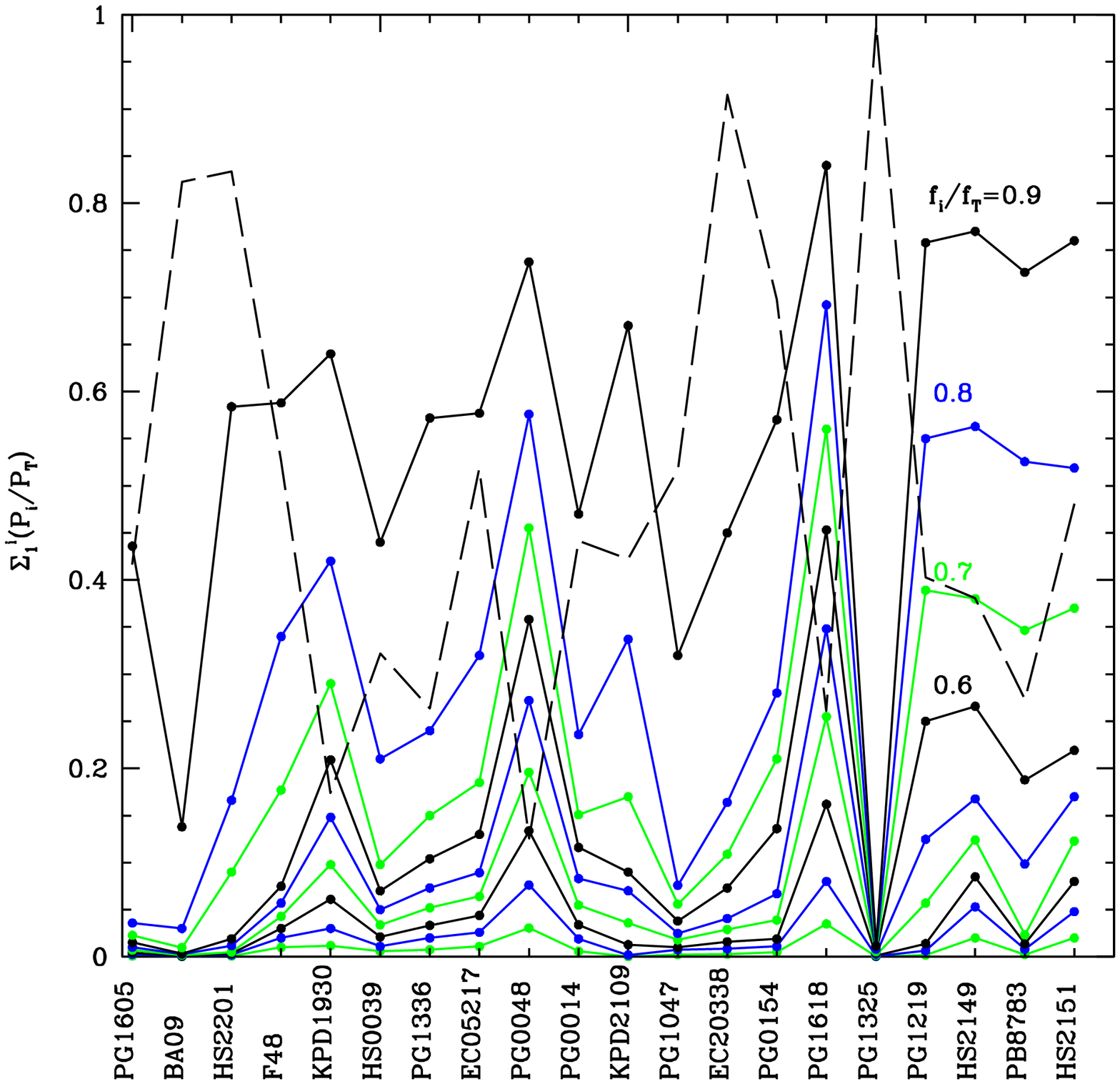,width=3.7in}}
\caption{Distribution of amplitudes (left panel) and power (right
panel) for pulsators with at least five frequencies. The stellar
designations are provided at the bottom and are organized by
increasing gravity similar. Each colored line indicates a
decreasing fraction of pulsation frequencies (from top to bottom),
with the first four lines labeled. The dashed line indicates the
fractional amplitude or power of the highest amplitude frequency
in each star.\label{fig13}}
\end{figure*}

\begin{figure}
 \centerline{\psfig{figure=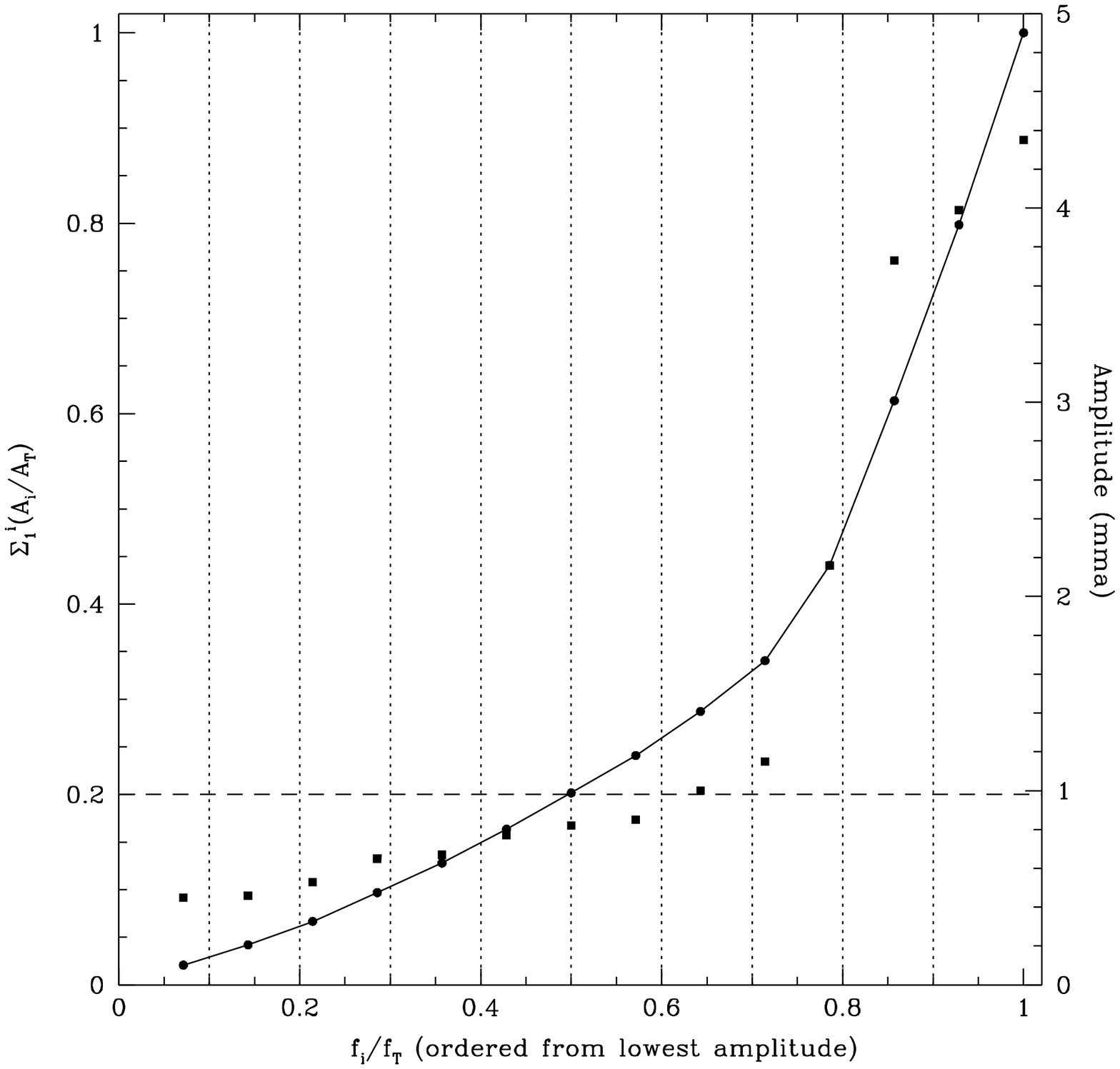,width=\textwidth}}
\caption{Distribution of amplitudes for HS~0039. The frequencies are
organized from lowest to highest amplitude with the amplitudes indicated
by filled squares (on the scale of the right-hand axis). The cumulative
fractional amplitudes are indicated by filled circles and connected by a
solid line. The vertical dotted lines are locate at 10\% intervals to 
indicate the points in Fig.~\ref{fig13} and the horizontal dashed line
(on the scale of the left-hand axis) indicates the
fractional amplitude of the highest amplitude frequency.\label{fig13a}}
\end{figure}

A more quantitative approach is shown in Fig.~\ref{fig13},
where we analyse the distribution of amplitudes and power in each
star.  For each
star, the frequencies were sorted by increasing amplitude and the
cumulative distribution of amplitudes (or power) was computed at
intervals of 10\% of the total. This process is shown
in Fig.~\ref{fig13a} for the star HS~0039. The 14 frequencies from
Table~\ref{tab03} are shown as squares with their amplitudes given by
the right-hand Y-axis. The circles indicate the cumulative fractional 
amplitude for each frequency, connected by the solid line, with their
scale given by the left-hand Y-axis. The points
for HS~0039 in the left panel of Fig.~\ref{fig13} correspond to
the locations in Fig.~\ref{fig13a} where the
dotted lines intersect the solid line. The top solid line in Fig.~\ref{fig13}
connects the points for the 90$^{\rm th}$ percentile, the next line 
for the 80$^{\rm th}$ percentile, and so on. The left panel
of Fig.~\ref{fig13} displays these points, or amplitude distribution, for
the 20 stars with more than five frequencies. Their designations are 
provided along the bottom in order of increasing
gravity.  Low values indicate that the top one or two
amplitudes contain a large fraction of the total amplitude (Fig.~\ref{fig13a}
would show a sharply peaked line).  The
right panel displays the distribution of power which
we define as the sum of the
squares of the amplitudes; the power and the maximum amplitude
$A_{\rm max}$ are provided in Table \ref{tab06}.  On average, the
lowest-amplitude half of the frequencies contribute about 25\% of
the total amplitude and 9.6\% of the total power.

If the amplitudes or power were evenly distributed (equal-height
amplitudes) then $f_i/f_T$ would equal $A_i/A_T$ (or $\sqrt{P_i/P_T}$). 
The corresponding distribution in Fig.~\ref{fig13a} would be a straight
diagonal line. There is a trend in Fig.~\ref{fig13} in that the contours 
are closest at low gravities and most
diffuse at high gravities. However, these trends are dominated by
four stars (two at each end) and are not representative of the
majority of the class. In the middle of each panel, there is a large
dispersion in values, with neighbouring stars transmitting most of
their power through one frequency, or distributing it more
equally.

In each panel, the dashed line indicates the fractional amplitude
or power emitted by the frequency with the highest amplitude. For
PG~1325, the one highest amplitude has 99\% of the total, while
for PG~0048 it is only $\sim 12$\%. The dashed line in Fig.~\ref{fig13a}
indicates that for HS~0039, the highest amplitude frequency has 20\% of
the combined amplitudes. While it is difficult to assign
significance to this plot as the number of pulsators is still
relatively low, it is suggestive that in
lower gravity stars the pulsation energy is channeled into
relatively few (or one) frequencies even though many frequencies
are available.

{\bf Multiplet structure:}
As the prototype for multiplet structure in pulsating white
dwarfs, PG~1159 showed that observational determination of the
modes can lead to tight constraints on the models
\citep{wing91}. With detailed studies of 23 pulsating sdBV stars,
it would be hoped that a similar star of this class would have
been detected. However, this has not been the case: multiplet
structure has been conspicuously absent, even from rich pulsators.
The only confirmed case of multiplets caused by rotational
splitting is Feige~48 \citep{me2,simon}, while another likely
candidate is BA09 \citep{andy}. % A model using splittings in
%PG~1605 was proposed by \citet{kaw99}; the model required a large
%rotation velocity which was later observed \citep{heber99}.
%However the model only incorporates 5 of 55 frequencies, making it
%somewhat unsatisfactory, and more recent observations ascribe the
%large velocity to pulsation rather than rotation \citep{kua}.
There are also marginal cases for rotationally induced multiplet
structure in HS~2201, PB~8783, PG~0014, PG~1047, and PG~1605 none of which
have been confirmed by additional observations. KPD~1930 and
PG~1336 are both known to be in short period binaries and
frequency splittings are commensurate with the binary period. An
initial interpretation did not attribute these to rotational
splitting, but rather to tidal effects induced by the companion
\citep{reed06c,reed06d}.

%From an observational point of view, these results are
%disconcerting because we should like to follow in the footsteps of
%white dwarf asteroseismologists who have used many types of
%observational constraints (including multiplet structure) to model
%detailed aspects of white dwarfs. 
A possible reason for the lack
of observed multiplet structure was proposed by \citet{kaw05}.
Their picture invokes sharp differential rotation in rapidly
spinning cores to modify the frequency spacings. Exceptions would
be for those stars in close binaries where rotations are tidally
locked.

Another multiplet pattern that has emerged lately is the
``Kawaler-scheme'' (Kawaler et al. 2006) a purely
mathematical formalism based on an asymptotic-like relationship
for the frequencies: 
$f(i,j)=f_o+i\times\delta +j\times\Delta$ where $i$ has integer
values, $j$ is limited to values of $-1,\,0,$ and $1$, $\delta$ is
usually a small spacing and $\Delta$ is usually a large spacing.
Their Table~3 indicates the significance of their predicted frequencies to
 those observed in several stars.  However, it is also known that the 
Kawaler-scheme does not fit several pulsators (including 
HS~0039) and so we merely make note of the scheme but await
a full report in a forthcoming paper.

% \begin{table}
% \centering
% \caption{Kawaler-scheme splittings found in sdBV stars. All frequencies
% and splittings reported in $\mu$Hz. (Reproduced
% directly from Kawaler et al. 2006).}
% \label{tab10}
% \begin{tabular}{lccccc}
% Star & $\log g$ & $f_o$ & $\delta$ & $\Delta$ & Significance \\ \hline
% PG~0014 & 5.8 & 5923 & 90.4 & 101.2 & 4.0$\sigma$ \\
% PG~1219 & 5.85 & 5812 & 60.5 & - & 3.7$\sigma$ \\
% PB~8783 & $>$5.6 & 7193 & 58.1 & 138.9 & 3.9$\sigma$ \\
% PG~1047 & 5.9 & 6310 & 55.9 & 186.5 & 4.2$\sigma$ \\
% PG~1336 & 5.7 & 4886 & 34.4 & 114.5 & 6.0$\sigma$ \\
% Feige~48 & 5.5 & 2642 & 13.9 & - & - \\ \hline
% \end{tabular}
% \end{table}

{\bf Frequency density:}
As discussed in \S 4.2, another tool at our disposal is the mode
density.  Although the mode density does not help to assign modes
to individual frequencies, we can set limits on the number of
degrees ($\ell$) per order ($n$) required to create the observed
frequency density from currently available models of
\citet[][hereafter CFB01]{char01} and of \citet{me2}.

\begin{figure}
 \centerline{\psfig{figure=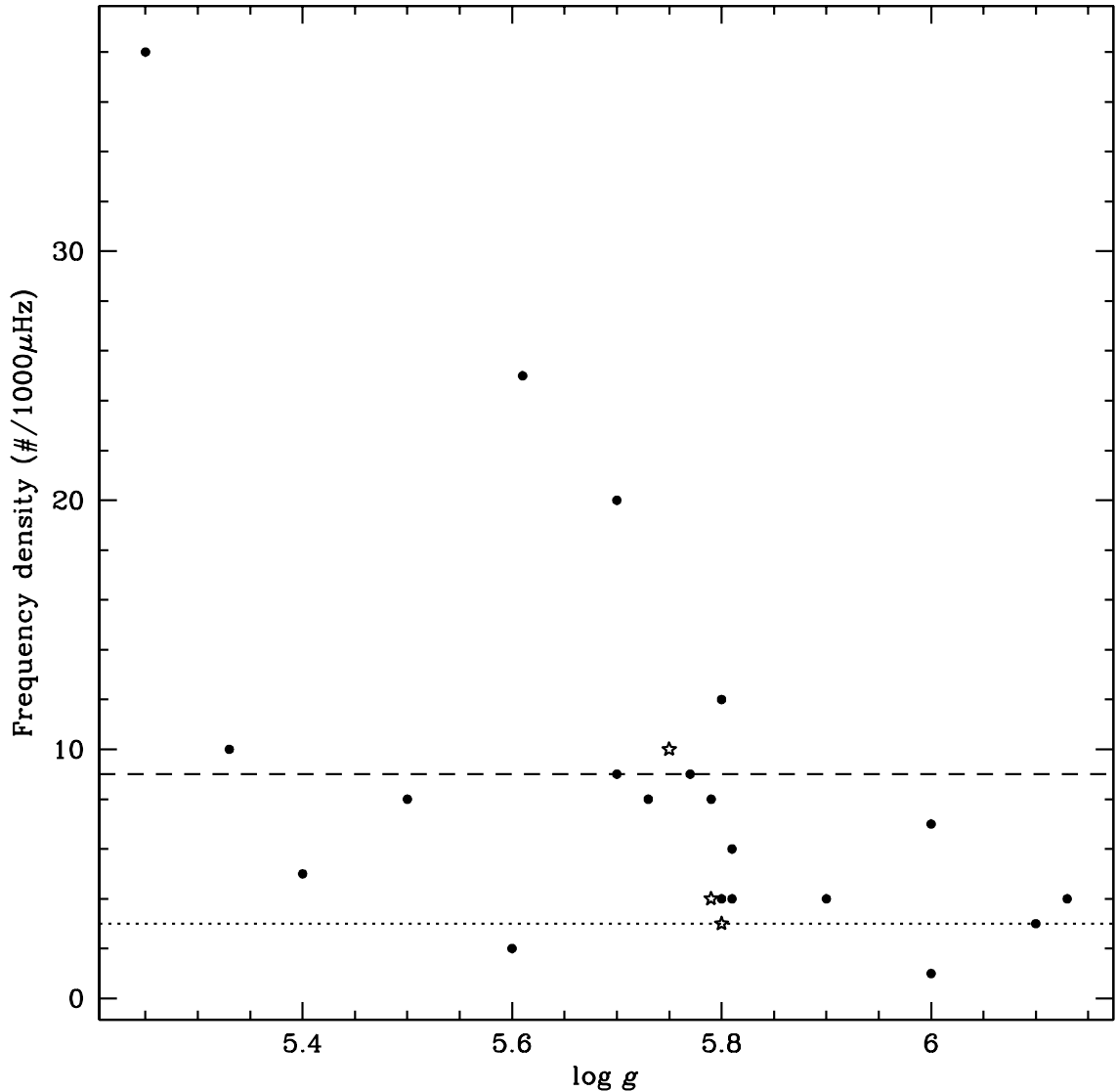,width=\textwidth}}
\caption{Mode density of pulsators organized by gravity. The
dotted line indicates the limit of 3 frequencies per 1000~$\mu$Hz
(all $m=0$ frequencies) and the dashed line indicates the limit of
9 frequencies per 1000~$\mu$Hz (all possible $2\ell +1$ $m$
values) assuming $\ell\leq 2$.\label{fig14}}
\end{figure}

Figure~\ref{fig14} shows the mode
density, plotted against $\log g$, with the dotted line indicating
3 frequencies per 1000~$\mu$Hz (all $\ell\leq 2,m=0$ modes) and
the dashed line indicating 9 frequencies per 1000~$\mu$Hz (all
possible $2\ell +1$ $m$ modes). Starred points indicate those
pulsators for which we have inferred the $\log g$ values. Less
than 20\% of sdBV stars have mode densities below 3 per
1000~$\mu$Hz, while more than 25\% have mode densities too high to
be reconciled with $\ell \leq 2$ modes even if all possible $m$
values are used. It therefore seems reasonable to conclude that
$\ell \geq 3$ modes must be excited.

\begin{figure}
 \centerline{\psfig{figure=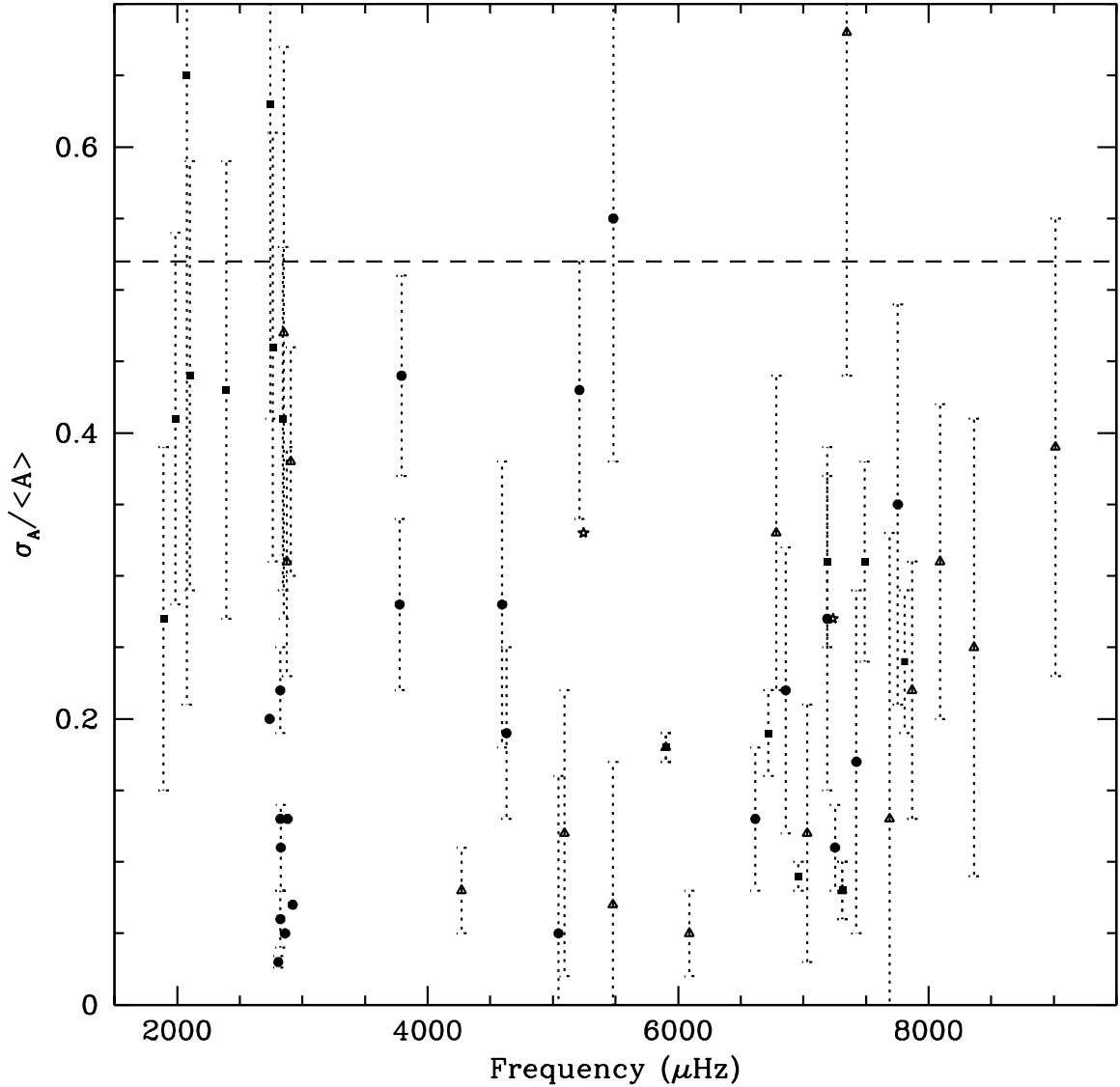,width=\textwidth}}
\caption{$\sigma (A)/\langle A\rangle$ of individual frequencies.
Triangles (squares) indicate phase-stable (-not-stable) frequencies,
circles indicate frequencies with no phase information or 
indeterminate phase stability and stars indicate the PG~0048 frequencies
which appear stochastic in nature. The dashed line is the JCD01 stochastic
value. \label{fig17}}
\end{figure}

{\bf Amplitude variability:}
A criterion outlined in Christensen-Dalsgaard
et al (2001; hereafter JCD01) and previously applied to several
sdBV stars (Pereira \& Lopes 2005; Zhou et al. 2006; Reed et al.
2006a, 2007) is to compare the average amplitude $\langle
A\rangle$ to the standard deviation of the amplitudes, $\sigma
(A)$. For stochastically excited pulsations, this ratio should
have a value near 0.5. For all resolvable frequencies in our
target stars, we have calculated both parameters and their ratios
which are given in columns 3 to 5 of Table~\ref{tab07} and plotted in
Fig.~\ref{fig17}. Columns 6 and
7 of Table~\ref{tab07} list the maximum
and minimum observed amplitudes and column 8 gives the maximum time-scale
over which the observations are sensitive to amplitude variations. In
the Figure, triangles (squares) indicate frequencies known to have stable 
(not-stable) phases, circles indicate frequencies with ambiguous or no
phase information and stars indicate the PG~0048 frequencies, which
are known to have stochastic-like properties (Reed et al. 2007). Stochastic
oscillations do not have stable pulsation phases and so 
frequencies with stable
phases should be driven rather than stochastically excited. Amplitudes
for HS~2201 came from Silvotti et al. (2002a) and no errors were published,
so no errorbars appear in the Figure.

Just like their average amplitudes and frequency density, the only
conclusion we can draw from Fig.~\ref{fig17} and Table~\ref{tab07}
is sdBV stars show a large variety of amplitude variations. Of course it
is known that many classes of pulsators show large amplitude differences
and that correlations between excitation rates and amplitudes are weak
at best, but these results indicate that there is no clear separation
in amplitude variability for phase-stable and unstable frequencies. As
such, it follows that the JCD01 criterion is likely not applicable for
these stars.
One last 
note is that our calculations do not include frequencies that have
been observed only one time (as in PG~1219 and Feige~48) or stars
for which we do not have the data. Both EC~14026 and EC~20338 are
reported to have sufficient amplitude variability that frequencies
completely disappear between observing seasons (Kilkenny et al.
2006b). 

% Table 8
\begin{table*}
\centering
\caption{Mean amplitude, standard deviation, and their ratio for
readily resolvable frequencies. Columns 6 through 8 provide the maximum
and minimum measured amplitudes and the longest time covered by the
observations. Except for HS~2201,
all amplitudes and standard deviations have been calculated
specifically for this paper from data we have. HS~2201 amplitudes
are directly from Silvotti et al. 2002a and no error estimates were
provided. Note that all frequencies
are in $\mu$Hz and all amplitudes are in mma.  \label{tab07}}
\begin{tabular}{lccccccc}
Star & Frequency & $\langle A\rangle$ & $\sigma (A)$ &$\sigma (A)/\langle A\rangle$
& $A_{max}$ & $A_{min}$ & Timescale \\ \hline
BA09 & 2807 & 45.76 & 1.36 & $0.03\pm 0.004$ & 52.89 & 44.30 & 53 days  \\
     & 2823 & 10.30 & 2.15 & $0.21\pm 0.03$ & 20.68 & 8.20 & 53 days \\
     & 2824 & 14.19 & 0.68 & $0.05\pm 0.02$ & 15.2 & 11.50 & 53 days \\
     & 2827 & 3.60 & 0.24 & $0.07\pm 0.03$ & 4.86 & 2.4 & 53 days \\
     & 3776 & 1.58 & 0.57 & $0.36\pm 0.06$ & 4.35 & 1.4 & 53 days \\
     & 3791 & 1.26 & 0.50 & $0.39\pm 0.07$ & 2.39 & 0.14 & 53 days  \\ \hline
EC~05217 & 4595 & 3.04 & 0.79 & $0.26\pm 0.10$ & 4.59 & 2.74 & 9 days   \\
         & 4629 & 4.32 & 0.73 & $0.17\pm 0.06$ & 5.16 & 3.05 & 9 days \\ \hline
Feige~48 & 2851 & 3.83 & 2.85 & $0.74\pm 0.20$ & 9.46 & 1.31 & 8 years \\
         & 2877 & 5.80 & 2.14 & $0.37\pm 0.08$ & 10.4 & 1.89 & 8 years \\
         & 2906 & 4.10 & 1.44  & $0.35\pm 0.08$ & 5.28 & 0.58 & 8 years \\ \hline
HS~0039 & 4271 & 4.45 & 0.17 & $0.04\pm 0.03$ & 5.03 & 3.82 & 30 days \\
        & 5482 & 2.35 & 1.30 & $0.56\pm 0.17$ & 4.44 & 0.12 & 30 days \\
        & 7348 & 1.09 & 0.77 & $0.70\pm 0.24$ & 3.17 & 0.37 & 30 days \\ \hline
HS~0444 & 5903 & 2.59 & 0.44 & $0.17\pm 0.01$ & 3.16 & 1.76 & 30 days \\
        & 7312 & 11.04 & 0.61 & $0.06\pm 0.02$ & 12.39 & 10.11 & 30 days \\ \hline
HS~1824 & 7190 & 3.01 & 0.92 & $0.30\pm 0.06$ & 5.26 & 0.87 & 47 days \\ \hline
HS~2151 & 6616 & 3.89 & 0.47 & $0.12\pm 0.05$ & 5.14 & 3.36 & 23 days \\
        & 6859 & 1.73 & 0.42 & $0.24\pm 0.10$ & 2.45 & 1.31 & 23 days \\
        & 7424 & 1.19 & 0.22 & $0.18\pm 0.12$ & 1.66 & 0.95 & 23 days \\ \hline
HS~2201 & 2738 & 0.39$\dagger$ & 0.08 & 0.20 & 0.48 & 0.34 & 1 year \\
        & 2824 & 4.85$\dagger$ & 0.54 & 0.13 & 5.65 & 4.23 & 1 year \\
        & 2861 & 10.31$\dagger$ & 0.54 & 0.05 & 10.88 & 9.77 & 1 year \\
        & 2881 & 1.16$\dagger$ & 0.16 & 0.13 & 1.34 & 1.00 & 1 year \\
        & 2922 & 0.6$\dagger$ & 0.04 & 0.07 & 0.64 & 0.56 & 1 year \\ \hline
KPD~2109 & 5045 & 2.64 & 1.05 & $0.40\pm 0.11$ & 4.80 & 0.59 & 32 days \\
         & 5093 & 6.45 & 0.75 & $0.12\pm 0.10$ & 8.09 & 5.21 & 32 days \\
         & 5212 & 1.65 & 0.54 & $0.32\pm 0.09$ & 3.14 & 0.31 & 32 days \\
         & 5481 & 6.21 & 0.45 & $0.07\pm 0.10$ & 7.41 & 4.96 & 32 days \\ \hline
PB~8783 & 7870 & 1.68 & 0.27 & $0.16\pm 0.09$ & 2.4 & 1.4 & 14 days \\
        & 8092 & 1.19 & 0.24 & $0.20\pm 0.11$ & 1.6 & 0.5 & 14 days \\ \hline
PG~0048 & 5245 & 1.74 & 0.58 & $0.33\pm 0.09$ & 2.44 & 0.93 & 15 days \\
        & 7237 & 1.45 & 0.40 & $0.25\pm 0.10$ & 2.13 & 0.27 & 15 days \\ \hline
PG~0154 & 6090 & 9.57  & 0.21 & $0.02\pm 0.03$ & 10.28 & 8.97 & 8 days \\
        & 6785 & 3.75  & 1.16 & $0.31\pm 0.11$ & 5.31 & 1.50 & 8 days \\
        & 7032 & 3.57  & 0.83 & $0.23\pm 0.09$ & 5.05 & 1.33 & 8 days \\
        & 7688 & 2.61  & 0.28 & $0.11\pm 0.20$ & 3.28 & 0.84 & 8 days \\
        & 8362 & 1.12  & 0.22 & $0.19\pm 0.16$ & 1.86 & 0.13 & 8 days \\
        & 9015 & 1.03  & 0.21 & $0.21\pm 0.16$ & 2.53 & 0.51 & 8 days \\ \hline
PG~1219 & 6722 & 3.35 & 0.49 & $0.15\pm 0.03$ & 4.72 & 2.15 & 4 years \\
        & 6961 & 7.00 & 0.46 & $0.07\pm 0.01$ & 8.06 & 5.73 & 4 years \\
        & 7490 & 5.19 & 1.74 & $0.33\pm 0.07$ & 8.48 & 2.94 & 4 years \\
        & 7808 & 6.35 & 1.65 & $0.26\pm 0.05$ & 9.85 & 4.44 & 4 years \\ \hline
PG~1325 & 7253 & 23.21 & 2.12 & $0.09\pm 0.03$ & 26.04 & 18.58 & 14 days \\ \hline
\end{tabular}
\end{table*}

% Table 8
\begin{table*}
\centering
\caption{Table 7 continued}
\begin{tabular}{lccccccc}
Star & Frequency & $\langle A\rangle$ & $\sigma (A)$ &$\sigma (A)/\langle A\rangle$
& $A_{max}$ & $A_{min}$ & Timescale\\ \hline
PG~1605 & 1891 & 9.6683 & 3.38 & $0.35\pm 0.12$ & 16.30 & 7.80 & 7 years \\
        & 1986 & 12.06 & 4.62 & $0.38\pm 13$ & 14.77 & 2.0 & 7 years \\
        & 2076 & 24.51 & 23.74 & $0.97\pm 0.44$ & 56.92 & 8.4 & 7 years \\
        & 2102 & 29.24 & 12.32 & $0.42\pm 0.15$ & 48.9 & 13.4 & 7 years \\
        & 2392 & 3.52 & 1.54 & $0.44\pm 0.16$ & 6.65 & 2.2 & 7 years \\
        & 2743 & 14.61 & 8.48 & $0.58\pm 0.22$ & 29.0 & 5.02 & 7 years \\
        & 2763 & 6.80 & 2.88 & $0.42\pm 0.15$ & 10.89 & 2.5 & 7 years \\
        & 2845 & 4.87 & 1.71 & $0.35\pm 0.12$ & 7.1 & 2.0 & 7 years \\ \hline
        % & 4062 & 3.81 & 2.52 & 0.64 & 6.87 & 1.4  \\
        % & 4152 & 3.11 & 1.65 & 0.53 & 5.51 & 0.85  \\
        % & 4178 & 2.73 & 1.1 & 0.4 & 2.0 & 0.23  \\ \hline
PG~1618B & 7191 & 2.13 & 0.44 & $0.21\pm 0.12$ & 3.6 & 1.14 & 45 days  \\
         & 7755 & 1.75 & 0.50 & $0.28\pm 0.14$ & 3.50 & 0.76 & 45 days \\ \hline
\end{tabular}
\end{table*}

\begin{figure*}
 \centerline{\psfig{figure=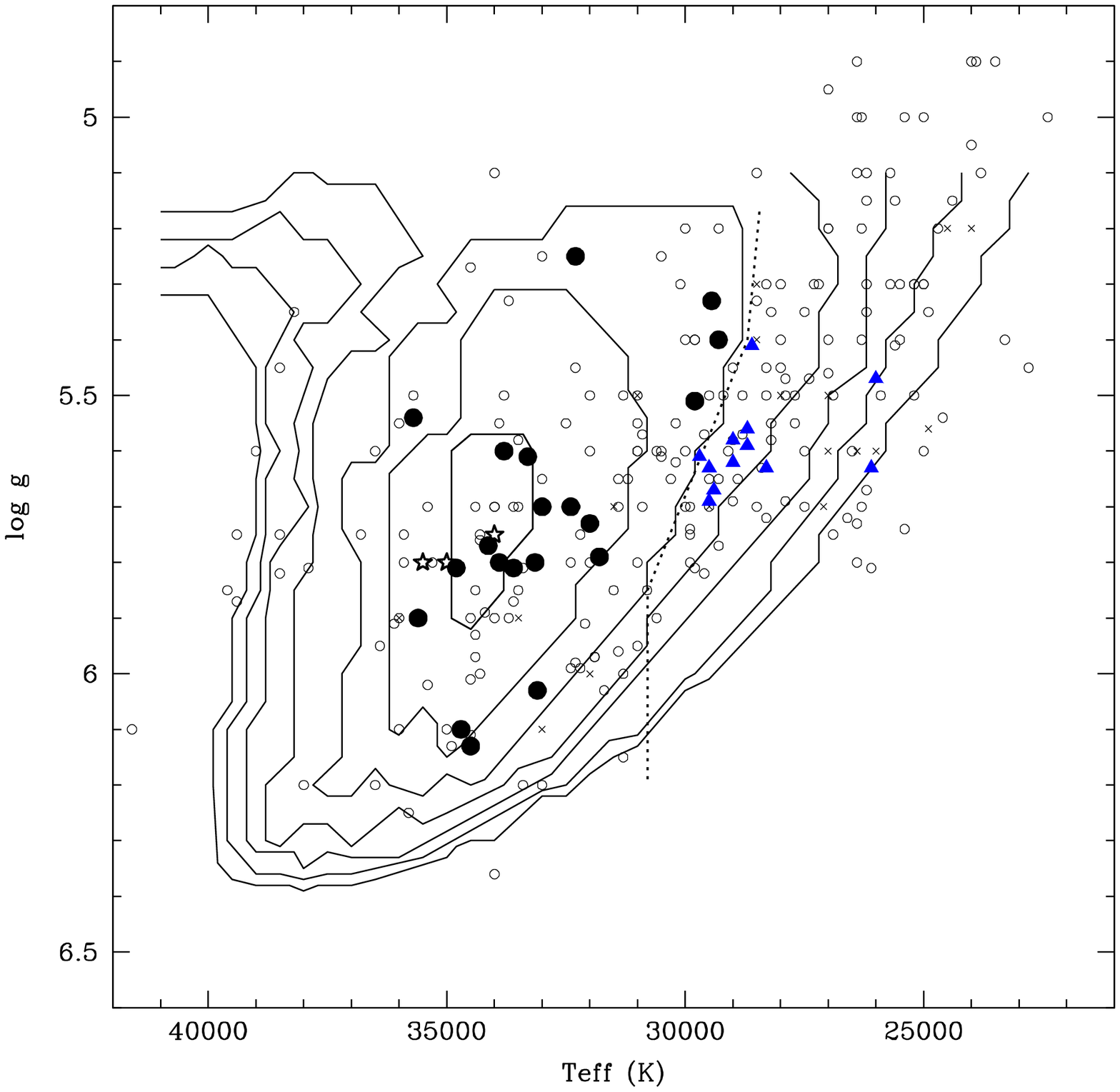,width=3.7in}\psfig{figure=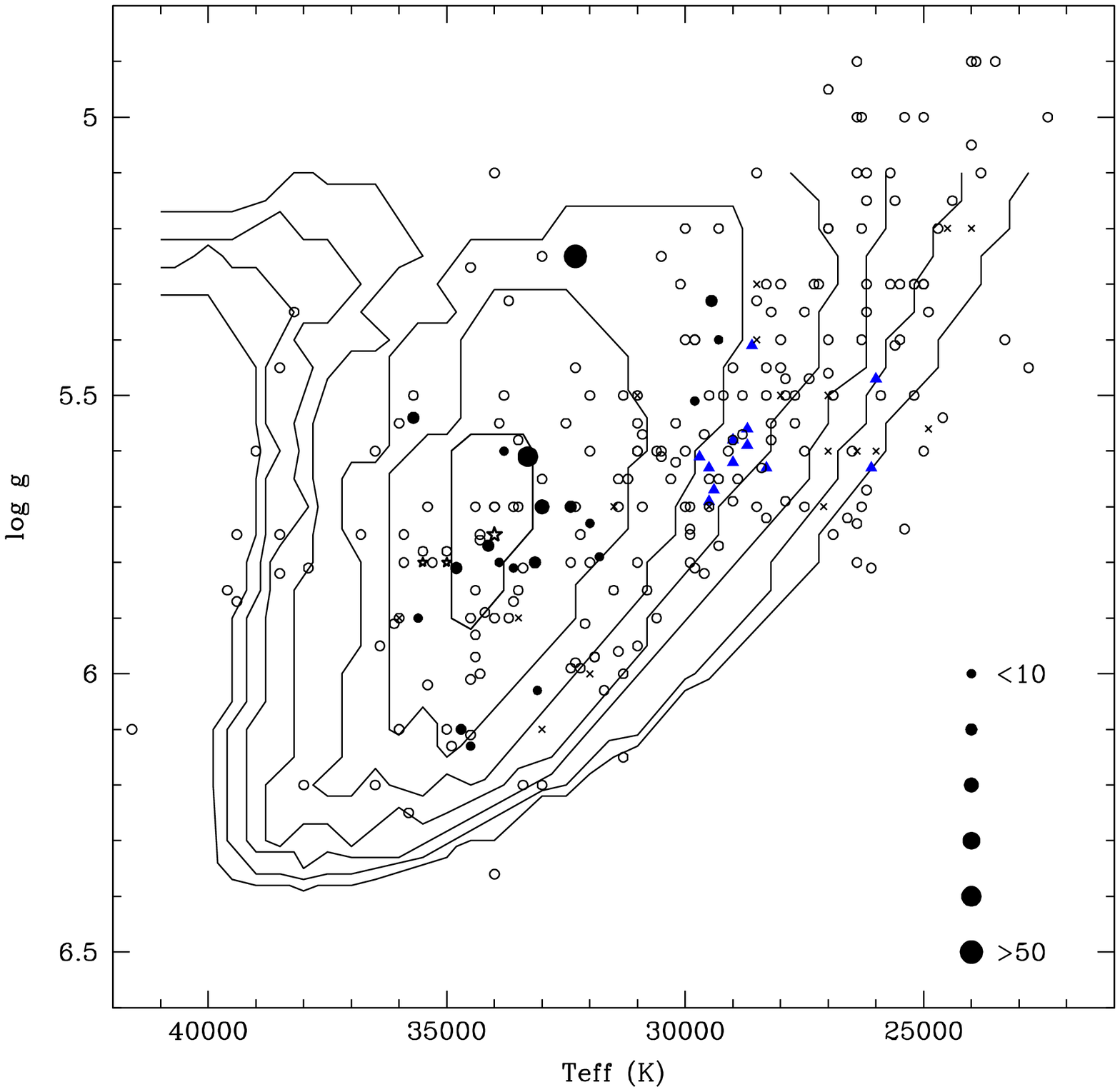,width=3.7in}}
\caption{Comparison of observed sdB stars to pulsational
instability contours (adapted from CFB01). Open circles are
non-pulsating sdB stars, filled circles are EC~14026-type
pulsators, and (blue) triangles are PG~1716-type pulsators. Left
panel includes a dashed line indicating the inferred separation
between EC~14026 and PG~1716-type pulsators and in the right panel
the dot size corresponds to the number of total pulsation
frequencies detected.\label{fig15}}
\end{figure*}

\begin{figure*}
 \centerline{\psfig{figure=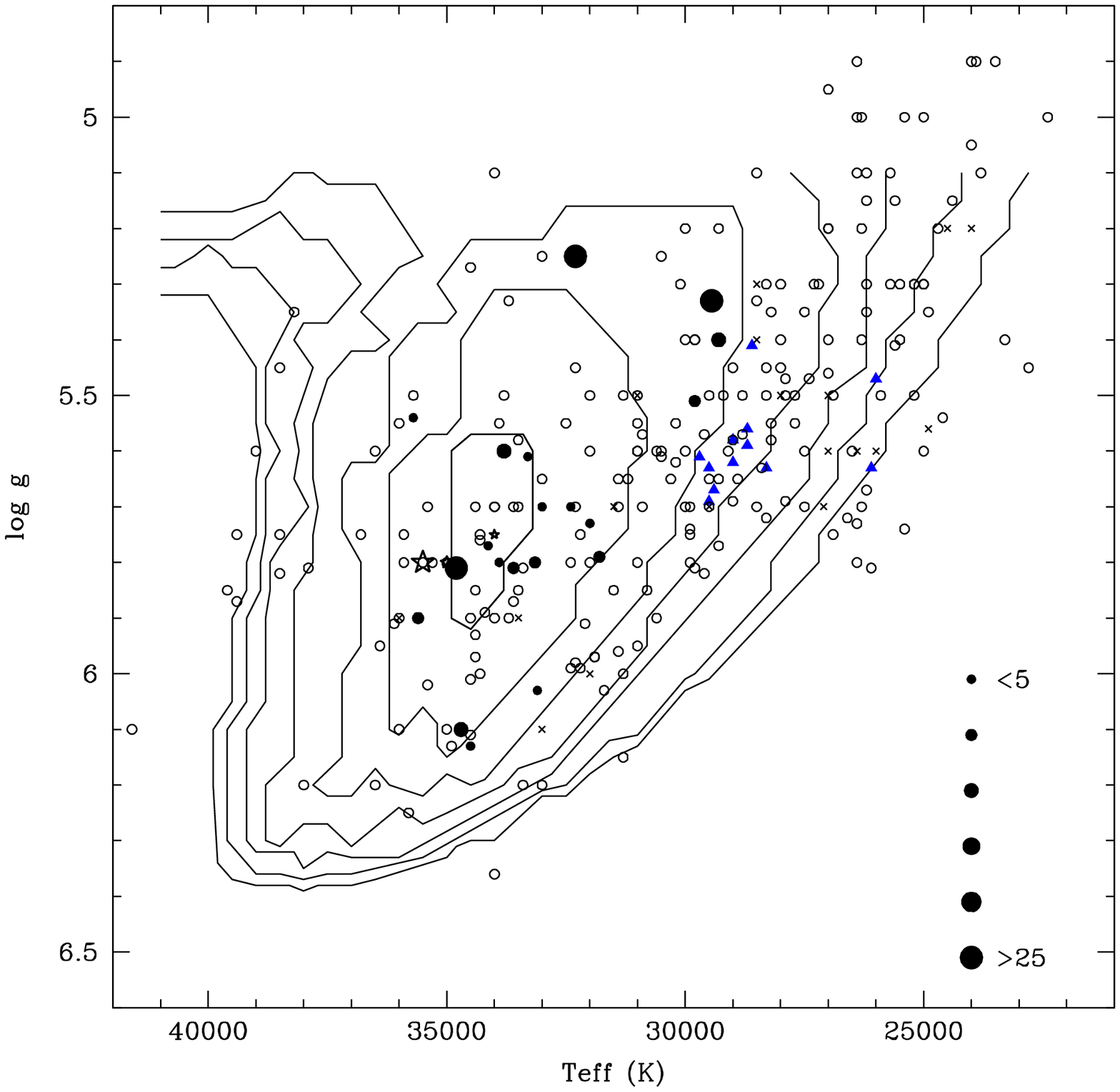,width=3.7in}\psfig{figure=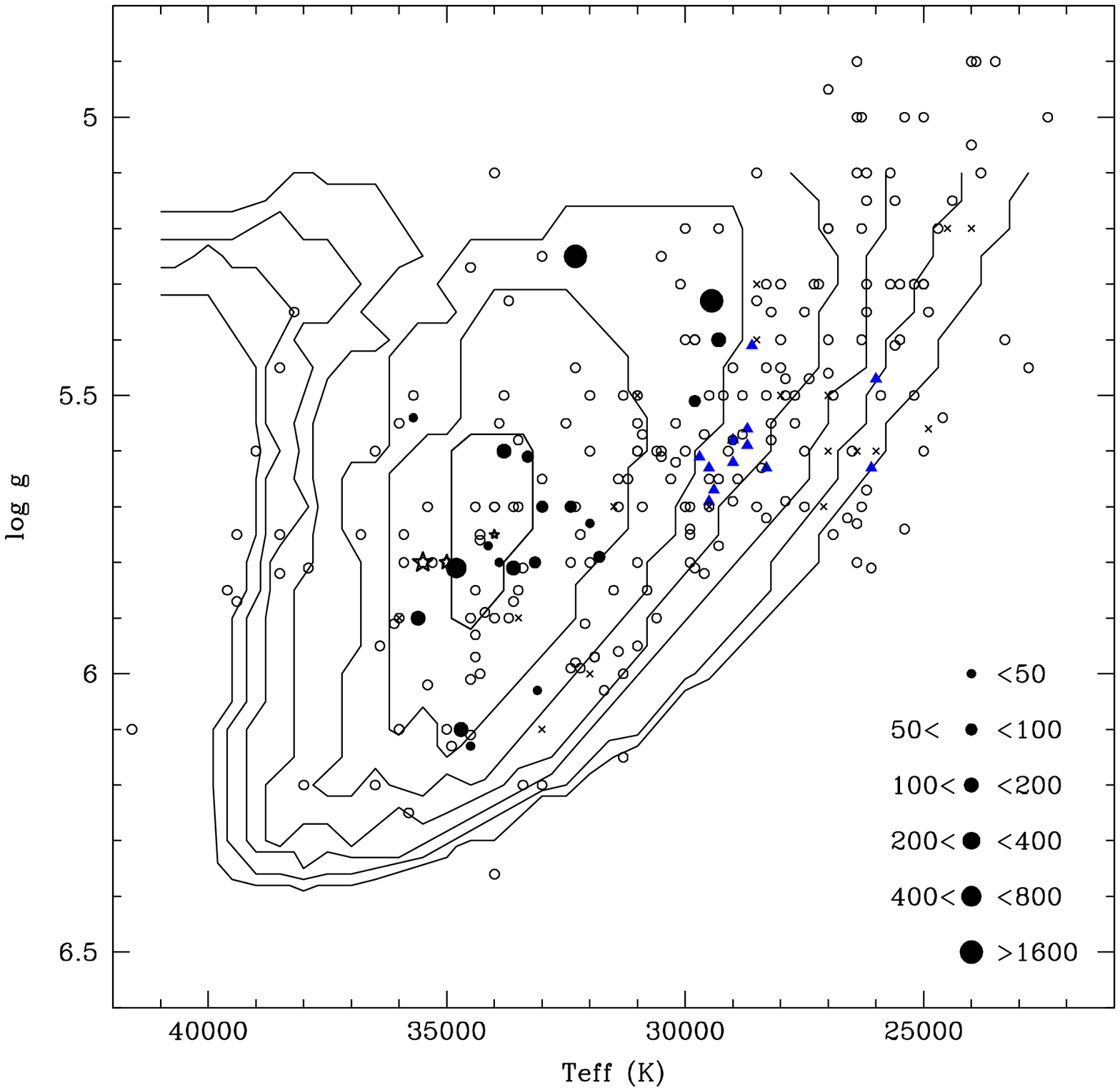,width=3.7in}}
\caption{Same as in Fig.~\ref{fig15} except that dot sizes for
pulsators represent $A_{max}$ (left panel) and total pulsation
power (right panel).\label{fig16}}
\end{figure*}

{\bf Comparison with theoretical instability contours:}
We can make a direct comparison between the
instability zone of the second-generation pulsation models
of \citet{char01} and observed
pulsation properties. Figures~\ref{fig15} and \ref{fig16} are used for
these discussions.
In all the plots, filled circles are the sdBV stars
of this study (stars indicate EC~20338, PG~0048, and
PG~0154\footnote{We have inferred $\log g$ from the shortest
pulsation frequency and $T_{\rm eff}$ from $B-V$ and $J-K_s$
colours for EC~20338, PG~0048, and PG~0154. These are very crude
estimates and should be considered as such.}), open circles are
sdB stars (and inferred to be non-pulsators) from the
Hamburg-Schmidt survey \citep{edel03}, Moehler et al (1990),
and Saffer et al. (1994), and filled (blue) triangles are
PG~1716-type pulsators (Green et al. 2003). The contours are reproduced
from CFB01
with the outside contour representing one unstable $\ell=0$
frequency with each interior contour representing an additional
unstable $\ell=0$ frequency up to $N=7$.

In the left panel of Fig.~\ref{fig15}, we determine how
discriminating the instability zone is by examining the
ratio of pulsators to non-pulsators within each contour.
However, we need to have an exception for the red (cool)
edge of the instability zone. There
is an indication that sdB stars may
\emph{switch} from EC~14026- to PG~1716-type pulsators in this
region. As such, this region (separated with a dotted line) should be
excluded; particularly since it is stated that ``all cool sdB
stars of low gravity may be PG~1716 pulsators'' (Fontaine et al.
2006). %Though we do not have a complete sample of non-pulsating
Working from the
inside ($N= 7$) contour outward, the fraction of pulsating to
non-pulsating sdB stars is: $N\geq 7$, 50\%; $N\geq 6$, 25\%;
$N\geq 5$, 25\%; $N\geq4$, 22\%; $N\geq 3$, 21\%; $N\geq 2$, 20\%;
$N\geq 1$, 20\% and the fraction of \emph{all} sdB stars within
the contours is $N\geq 7$, 12\%; $N\geq 6$, 52\%; $N\geq 5$, 80\%,
$N\geq4$, 90\%; $N\geq 3$, 92\%; $N\geq 2$, 95\%; $N\geq 1$, 95\%;
$N<1$, 100\%. There does appear to be a relationship between
the interior instability ($N=7$) contour and fraction of pulsators.
Yet outside of the first contour, the ratio only changes by 5\% and
all of the EC~14026-type pulsators are
within the first three contours. Yet so are 80\% of \emph{all} sdB
stars.
So while the contours match where the pulsators are,
they also match where \emph{most} sdB stars with $T_{\rm
eff}\geq$30~000~K are. %Therefore, the instability contours do not

The remaining panels of Figs.~\ref{fig15} and \ref{fig16} show
the same points, but the size of the dots correlate with the
following observed properties; the number of pulsation frequencies
(right panel of Fig.~\ref{fig15}), the maximum amplitude
($A_{max}$), and total pulsation power.
$A_{max}$ and total pulsation power are related as
power is dominated by a few high-amplitude frequencies. Stars
such as PG~0048 with many, but low amplitude frequencies cannot
match the power of a single $A\geq 20$~mma frequency. We would
expect a correlation of the instability
contours with the \emph{number} of frequencies
detected, as each interior contour increases the number of
theoretically unstable frequencies, but such is not the case. From
\S 5.2, we know that it is not a detection issue as rich pulsators
occur with both low and high amplitudes. In
fact there seems to be no correlation whatsoever with the largest
points (largest number of frequencies, highest amplitude and most
pulsation power) occupying multiple regions of the $\log
g - T_{\rm eff}$ diagrams and similar results for the smallest
points. As such the group properties do not add observational support to 
the driving theory.

\section{Conclusions}

From extensive follow-up data acquired at MDM observatory, we are
confident that we have resolved the pulsation spectra of two
additional pulsating sdB stars. For HS~0039, we detect 10
additional frequencies bringing the total to 14 and for HS~0444,
we confirm the two frequencies of the discovery data and detect an
additional low amplitude frequency. We have also noted that while
the amplitudes and phases of HS~0444 appear very steady over the
duration of our observations, those in HS~0039 did not, but
rather have varied considerably. This is illustrative of the
variety observed in sdBV stars where some stars can have very
simple and/or stable pulsation spectra while others can be quite
rich, with tens of frequencies that may change amplitudes on short
time scales.

Since the discovery of sdBV stars in 1996, more than 23 of the 34
known EC~14026-type pulsators have received follow-up
observations. We have examined these 23 stars for which extended
timebase (and often multisite) observations have been acquired.
% The most remarkable feature we find is the large amount of variety
% detected in their pulsations. Stars at the same gravity and
% temperature (within the errors) can have few or many pulsation
% frequencies with amplitudes that are large or small, stable or
% unstable. Seemingly stochastically excited frequencies can reside
% along side stable frequencies, be all that are observed, or absent
% altogether.
% 
We have searched for trends in frequency groupings, the number of
high (H) amplitude to total (T) frequencies, and frequency density as a
function of gravity and note that the only trend seems to be a
weak relationship between the H/T ratio and gravity: The lowest
gravity pulsators have the smallest H/T ratio (a few very high
amplitude frequencies) while the highest gravity pulsators have
H/T$\approx 1$ (relatively even amplitudes distributed amongst the
observed frequencies).

We examined amplitude stability, which has been used to infer stochastically
excited oscillations in other variable classes and previously applied
to sdB pulsators. And while many pulsation frequencies fit the JCD01
value, the distribution between those and 
amplitude-stable and presumably driven is relatively smooth with no clear
separations. However, phase-stable frequencies, some with quite large
variability, preclude them from being stochastically excited and so
the simplest conclusion is that the JCD01 criteria is not appropriate
for sdBV stars.

We compared pulsators to theoretical instability contours to
search for relationships that correlate with current driving
theory. While there is a weak concentration of pulsators precisely
where expected, these same contours include more than 80\% of
\emph{all} sdB stars in our sample with $T_{\rm eff}\geq
30\,000$~K (an inferred cut-off for PG~1716-type pulsators) and
there is no correlation between the contours and the number of
pulsation frequencies, nor their maximum amplitude or total
pulsation power. It would be nice if we could note that any
correlation is canceled by stars with lower gravities requiring
less energy to drive higher amplitudes which results in more
frequencies being detected, but this is also not the case as the
amplitudes of neighboring stars \emph{anywhere} in the $\log
g - T_{\rm eff}$ plane can have amplitudes that differ by more
than an order of magnitude.

In the end, what we have uncovered in our examination of sdBV
stars is the large variety they encompass. Subdwarf B pulsators
seem to exist at \emph{all} temperatures and gravities (though we
only examined the EC~14026-type pulsators at $T_{\rm eff}\geq
30\,000$~K) that sdB stars do and at roughly the same
concentrations. Regardless of temperature or gravity (and thus
evolutionary age and/or total mass and/or envelope mass and/or
metallicity) pulsators can have relatively high ($\geq 20$~mma) or
low ($\leq 2$~mma), stable or unstable amplitudes and the frequency
density can be quite high or very low. % (HS~1824 has only 1
%frequency) and they are nearly as likely to match stochastic
%criteria or the Kawaler-scheme which invoke alternate driving
%mechanisms, as the theoretically proposed $\kappa$ mechanism. We
%We look forward to the Charpinet et al. third-generation models,
%which we expect will incorporate more physics and place tighter
%constraints on instability, as well as efforts in alternate
%directions, such as those by Kawaler et al. who have investigated
%rapid and differential interior rotation and devised a purely
%mathematical formalism of frequency matching. 
We hope that our
investigation of these 23 well-studied sdBV stars will be useful
and bring new insight to their theoretical perspectives.
Additionally, we also hope that this study will be useful for
observers pursuing multicolour photometry and time-series
spectroscopy. We anticipate that such works will be required for
an overall understanding of sdB pulsations, yet these more
detailed observations require that the pulsation characteristics
of the star be known beforehand. So even though 23 of 34 known
sdBV stars have received follow-up observations, there is still a
long way to go both observationally and theoretically.

ACKNOWLEDGMENTS: We would like to thank the MDM and McDonald
observatory TACs for generous time allocations, without which this
work would not have been possible. We would also like to thank
Dave Mills for his help with the Linux camera drivers, Darragh
O'Donoghue, Chris Koen, Dave Kilkenny and Andrzej Baran for kindly
sharing their data. We also thank the many observers who have
provided support for our campaigns, particularly those at Lulin
and Suhora observatories who have helped on many campaigns.
Support for DMT came in part from funds provided by the Ohio State
University Department of Astronomy. This material is based in part
upon work supported by the National Science Foundation under Grant
Number AST007480. Any opinions, findings, and conclusions or
recommendations expressed in this material are those of the
author(s) and do not necessarily reflect the views of the National
Science Foundation.

\end{document}